\documentclass[12pt]{article}
\usepackage{float}
\usepackage{multicol}
\usepackage{mathtools}
\usepackage{listings}
\usepackage{color}
\usepackage{pdfpages}
\usepackage{amssymb}
\usepackage{tabularx}
\usepackage{bm}
\usepackage{url}
\usepackage{authblk}

\usepackage[table,xcdraw]{xcolor}

\usepackage[yyyymmdd,hhmmss]{datetime}

\ifnum\currenthour<5
  \advance\currenthour+20
  \ifnum\day=1
    \testifleapyear\year
    \ifnum\month=1 \month13\advance\year-1\fi
    \advance\month-1
    \day\ifcase\month\or31\or\ifDTleapyear 29\else 28\fi\or31\or30\or31\or30\or31\or
                                                           31\or30\or31\or30\or31\fi
  \else\advance\day-1\fi
\else\advance\currenthour-4\fi

\usepackage[round]{natbib}
\usepackage{amsmath,amssymb,amsthm,bm,enumerate,mathrsfs,mathtools}
\usepackage{latexsym,color,verbatim,multirow}
\usepackage[margin=1in]{geometry}
\usepackage{graphicx,hyperref}
\usepackage{caption}
\usepackage{footmisc}
\usepackage{lscape}
\usepackage{setspace}
\usepackage{newfloat}
\usepackage[para,online,flushleft]{threeparttable}
\DeclareFloatingEnvironment[name={Appendix Figure}]{suppfigure}
\usepackage{zwcommands}

\newcommand{\logit}{{\sf{logit}}}

\newcommand{\BrS}{{\tiny \sf{BrS}}}
\newcommand{\SSs}{{\tiny \sf{SS}}}

\usepackage{titlesec}

\titleformat{\paragraph}
{\normalfont\normalsize\bfseries}{\theparagraph}{1em}{}
\titlespacing*{\paragraph}
{0pt}{3.25ex plus 1ex minus .2ex}{1.5ex plus .2ex}

\usepackage{xcolor}
\hypersetup{
  colorlinks,
  linkcolor={red!50!black},
  citecolor={blue!50!black},
  urlcolor={blue!80!black}
}

  \usepackage{circuitikz}

\usepackage{booktabs}

\newcommand{\blind}{0} 
\begin{document}
	\def\spacingset#1{\renewcommand{\baselinestretch}%
		{#1}\small\normalsize} \spacingset{1}

\if0\blind
	{
		\title{\bf Regression Analysis of Dependent Binary Data for Estimating Disease Etiology from Case-Control Studies}
		\author[1,2]{Zhenke Wu}
		\author[1]{Irena Chen}
		\affil[1]{Department of Biostatistics, University of Michigan, Ann Arbor, MI 48109, USA; E-mail: {\tt zhenkewu@umich.edu}.}
		\affil[2]{Michigan Institute for Data Science, University of Michigan, Ann Arbor, MI 48109, USA.}
		\date{}
		\maketitle
} \fi

\if1\blind
	{
		\title{\bf Bayesian Regression Analysis for Estimating Disease Etiology}
		\author[]{}
		\date{}
		\maketitle
	} \fi

\vspace{-1cm}

\bigskip
\begin{abstract}


In large-scale disease etiology studies, epidemiologists often need to use multiple binary measures of unobserved causes of disease that are not perfectly sensitive or specific to estimate cause-specific case fractions, referred to as ``population etiologic fractions" (PEFs). Despite recent methodological advances, the scientific need of incorporating control data to estimate the effect of explanatory variables upon the PEFs, however, remains unmet. In this paper, we build on and extend nested partially-latent class model \citep[npLCMs,][]{wu2017nested} to a general framework for etiology regression analysis in case-control studies. Data from controls provide requisite information about measurement specificities and covariations, which is used to correctly assign cause-specific probabilities for each case given her measurements. We estimate the distribution of the controls' diagnostic measures given the covariates via a separate regression model and \textit{a priori} encourage simpler conditional dependence structures. We use Markov chain Monte Carlo for posterior inference of the PEF functions, cases' latent classes and the overall PEFs of policy interest. We illustrate the regression analysis with simulations and show less biased estimation and more valid inference of the overall PEFs than an npLCM analysis omitting covariates. An regression analysis of data from a childhood pneumonia study site reveals the dependence of pneumonia etiology upon season, age, disease severity and HIV status.

\end{abstract}
\noindent%
\newpage
\spacingset{1.45}

\newpage
\section{Introduction}
\label{sec::intro}

In epidemiologic studies of disease etiology, one important scientific goal is to assess the effect of explanatory variables upon disease etiology. Based on multiple binary non-gold-standard diagnostic measurements made on a list of putative causes with different error rates, this paper develops and demonstrates a regression analytic approach for drawing inference about the cause-specific fractions among the case population that depend on covariates. We illustrate the analytic needs raised by a study of pediatric pneumonia etiology.

Pneumonia is a clinical condition associated with infection of the lung tissue, which can be caused by more than 30 different species of microorganisms, including bacteria, viruses, mycobacteria and fungi \citep{scott2008pneumonia}. The Pneumonia Etiology Research for Child Health (PERCH) study is a seven-country case-control study of the etiology of severe and very severe pneumonia and has enrolled more than $4,000$ hospitalized children  under five years of age and more than $5,000$ healthy controls \citep{o2019aetiology}. The goal of the PERCH study is to estimate the population fractions of cases due to the pathogen causes, referred to as ``population etiologic fractions" (PEFs) and to assign cause-specific probabilities for each pneumonia child given her measurements, termed as ``individual etiologic fractions" (IEFs). The PERCH study also aims to understand the variation of the PEFs as a function of factors such as region, season, a child's age, disease severity, nutrition status and human immunodeficiency virus (HIV) status.

The cause of lung infection cannot, except in rare cases, be directly observed \citep{hammitt2017addressing}. The PERCH study tests the presence or absence of a list of pathogens using specimens in peripheral compartments including the blood, sputum, pleural fluid and nasopharyngeal (NP) cavity \citep{crawley2017standardization}. In this paper, we focus on two sources of imperfect measurements: ({\sf a}) NP Polymerase Chain Reaction (NPPCR) results from cases and controls that are not perfectly sensitive or specific, referred to as ``bronze-standard" (BrS) data; and  ({\sf b}) blood culture (BCX) results from cases only that are perfectly specific but lack sensitivity, referred to as ``silver-standard" (SS) data.

Valid inference about the population and individual etiologic fractions must address three salient characteristics of the measurements. First, tests lacking sensitivity such as NPPCR and BCX may miss true causative agent(s) which if unadjusted may underestimate the PEFs. Second, imperfect diagnostic specificities may result in the detection of multiple pathogens in NPPCR that may indicate asymptomatic carriage but not causes of pneumonia. Determining the primary causative agent(s) must use statistical controls. Third, multiple specimens are tested among the cases with only a subset available from the controls. Other large-scale disease etiology studies have raised similar analytic needs and challenges of integrating multiple sources of imperfect measurements of multiple pathogens to produce an accurate understanding of etiology \citep[e.g.,][]{saha2018causes, kotloff2013burden}.

To address the analytic needs, \cite{wu2016partially} introduced a \textit{partially-latent class model} (pLCM) as an extension to classical latent class models \citep[LCMs][]{lazarsfeld1950thelogical, Goodman1974} that uses case-control data to estimate the PEFs. This prior work shows the capacity of the multivariate specimen measurements to inform the distribution of unobserved, or ``latent" health status for an individual and the population. PLCM is a finite mixture model with $L+1$ components for multivariate binary data where a case observation is drawn from a mixture of $L$ components each representing a cause of disease, or ``disease class"; Controls have no infection in the lung hence are assumed drawn from an observed class. The pLCM is a \textit{semi-supervised} method for learning the unobserved classes, where the ``label" (cause of disease) is observed for only a subset of subjects. Let $I_i\in \{1, \ldots, L\}$ represent case $i$'s disease class which is categorically distributed with probabilities equal to the PEFs $\bpi = (\pi_1, \ldots, \pi_L)^\top$ in the $(L-1)$-dimensional simplex $\cS_{L-1}=\{\bpi: \sum_{\ell=1}^L \pi_\ell = 1, 0\leq \pi_\ell\leq 1\}$. A case class can represent a single- or multiple-pathogen cause of pneumonia, or pathogen causes not targeted by the assays which we refer to as ``{\sf Not Specified} ({\sf NoS})". PLCM uses a vector of $J$ response probabilities to specify the conditional distribution of the measurements in each class. PLCM is an example of \textit{restricted} LCMs \citep[RLCMs,][]{wu2019barlcm} which restrict how the response probabilities differ by class to reflect the scientific knowledge that causative pathogens are more likely to appear in the upper respiratory tract in a pneumonia child than a healthy control. In particular, each causative pathogen is assumed to be observed with a higher probability in case class $\ell$ (sensitivity or true positive rate, TPR) than among the controls; A non-causative pathogen is observed with the same probability as in the controls (1 - specificity or false positive rate). Under the pLCM, a higher observed marginal positive rate of pathogen $j$ among cases than controls indicates etiologic significance.

In a Bayesian framework, measurements of differing precisions can be optimally combined under a pLCM to generate stronger evidence about $\bpi$. The pLCM is partially-identified \citep{Jones2010,gu2019partial}. There exist two sets of values of a subset of model parameters (here the TPRs) that the likelihood function alone cannot distinguish even with infinite samples; Bounds on the parameters however are available \citep[e.g.,][Equation 6]{wu2016partially}. Informative prior distributions for the TPRs elicited from laboratory experts or estimated from vaccine probe studies for a subset of pathogens \citep{feikin2014use} can be readily incorporated to improve inference \citep{gustafson2015bayesian}. 


The pLCM assumes ``local independence" (LI) which means the BrS measurements are mutually independent given the class membership. This classical assumption is central to mixture models for multivariate data, because the estimation procedures essentially find the optimal partition of observations into subgroups so that the LI approximately holds in each subgroup. Deviations from LI, or ``local dependence" (LD) are testable using the control BrS data, which can be accounted for by an extension of pLCM, called \textit{nested partially-latent class model} \citep[npLCM,][]{wu2017nested}. In each class, the npLCM uses the classical LCM formulation that has the capacity to describe complex multivariate dependence among discrete data \citep{dunson2009nonparametric}. For example, it assumes the within-class correlations among NPPCR tests are induced by unobserved heterogeneity in subjects' propensities for pathogens colonizing the nasal cavities.  In particular, LD is induced in an npLCM by nesting $K$ latent subclasses within each class $\ell=0,1,...,L$, where subclasses respond with distinct vectors of probabilities. In a Bayesian framework with a prior that encourages few important subclasses, the npLCM reduces the bias in estimating $\bpi$, retains estimation efficiency and offers more valid inference under substantial deviation from LI.



Extensions to incorporate covariates in an npLCM are critical for two reasons. Firstly, covariates such as season, age, disease severity and HIV status may directly influence $\bpi$. Secondly, in an npLCM without covariates, the relative probability of assigning a case subject to class $\ell$ versus class $\ell'$ depends on the FPRs \citep{wu2016partially} which are estimable using the control data. However, the FPRs may vary by covariates which if not modeled will bias the assignment of cause-specific probabilities for each case subject. For example, pathogen {\sf A} found in a case's nasal cavity less likely indicates etiologic significance than a colonization during seasons with high asymptomatic carriage rates, and much more so when the same pathogen rarely appears in healthy subjects.

Adapting existing no-covariate methods to account for discrete covariates, one may perform a \textit{fully-stratified analysis} by fitting an npLCM to the case-control data in each covariate stratum. Like pLCM, the npLCM is partially-identified in each stratum \citep{wu2017nested}, necessitating multiple sets of \textit{independent} informative priors across multiple strata. There are two primary issues with this approach. First, sparsely-populated strata defined by many discrete covariates may lead to unstable PEF estimates. Second, it is often of policy interest to quantify the overall cause-specific disease burdens in a population. Let the overall PEFs $\bpi^\ast=(\pi^\ast_1, \ldots, \pi^\ast_L)^\top$ be the empirical average of the stratum-specific PEFs. Since the informative TPR priors are often elicited for a case population and rarely for each stratum, reusing independent prior distributions of the TPRs across all the strata will lead to overly-optimistic posterior uncertainty in $\bpi^\ast$, hampering policy decisions. 


Estimating disease etiology across discrete and continuous epidemiologic factors needs new methods in a general modeling framework. In this paper, we extend the npLCM to perform regression analysis in case-control disease etiology studies that ({\sf a}) incorporates controls to estimate the PEFs, ({\sf b}) specifies parsimonious functional dependence of $\bpi$ upon covariates such as additivity, and ({\sf c}) correctly assesses the posterior uncertainty of the PEF functions and the overall PEFs $\bpi^\ast$ by applying the TPR priors just once. 




The rest of the paper is organized as follows. Section \ref{sec:overview_noX} overviews the npLCM without covariates. Section \ref{sec:nplcm_withX} builds on the npLCM and makes the regression extension. We demonstrate the estimation of disease etiology regression functions $\pi_\ell(\cdot)$ through simulations in Section \ref{sec:simulation}; We also show superior inferential performance of the regression model in estimating the overall PEFs $\bpi^\ast$ relative to an analysis omitting the covariates. In Section \ref{sec:perch}, we characterize the effect of seasonality, age, HIV status upon the PEFs by applying the proposed npLCM regression model to the PERCH data. The paper concludes with a discussion.

\section{Overview of npLCMs without Covariates}
\label{sec:overview_noX}

Let binary BrS measurements $\bm{M}_{i} = (M_{i1},...,M_{iJ})^\top$ indicate the presence or absence of $J$ pathogens for subject $i=1, \ldots, N$. Let $Y_i$ indicate a case ($1$) or a control ($0$) subject. If $Y_i=1$, let $I_i \in \{1, \ldots, L\}$ represent case $i$'s unobserved disease class; Otherwise, let $I_i=0$ because a control subject's class is known (in PERCH, no lung infection).  In this paper, we simplify the presentation of  models by focusing on single-pathogen causes (hence $L=J$). The npLCM readily extends to $L>J$ for including additional pre-specified multi-pathogen and/or ``{\sf Not Specified}" ({\sf NoS}) causes \citep{wu2017nested}. 


The likelihood function for an npLCM has three components: ({\sf a}) PEFs or cause-specific case fractions: $\bpi = (\pi_{1}, \ldots, \pi_{L})^\top =\{\pi_{\ell} = \PP(I=\ell\mid Y=1), \ell=1, \ldots, L\} \in \cS_{L-1}$; ({\sf b}) $\bP_{1\ell}=\{\bP_{1\ell}(\bm{m})\}=\{\PP(\bM=\bm{m} \mid I=\ell, Y=1)\}$: a table of probabilities of making $J$ binary observations $\bM=\bm{m}$ in a case class $\ell\neq 0$; ({\sf c}) $\bP_0 =\{\bP_0(\bm{m})\}=\{ \PP(\bM=\bm{m} \mid I=0, Y = 0)\}$: the same probability table but for controls. Since cases' disease classes are unobserved, the distribution of  cases' measurements $\bP_1=\PP(\bM \mid Y=1)$ is a finite-mixture model with weights $\bpi$ for the $L$ disease classes: \(\bP_1= \sum_{\ell=1}^L\pi_{\ell}\bP_{1\ell}\). 


Models in this section differ by how $\bP_0$ and \{$\bP_{1\ell}$\} are specified; Regression models in Section \ref{sec:nplcm_withX} further incorporate covariate into the specifications ($\bpi$ as well). More specifically, the likelihood of an npLCM \citep{wu2017nested} is a product of case ($L_1$) and control ($L_0$) likelihood functions
\begin{align}
L = L_1\cdot L_0=\left\{\prod_{i:Y_i=1}\sum_{\ell=1}^L \pi_\ell\cdot \bP_{1\ell}(\bM_i; \bTheta,\bPsi,\bEta)\right\}\times  \prod_{i':Y_{i'}=0}\bP_0(\bM_{i'}; \bPsi,\bnu),\label{eq:no_x_lkd}
\end{align}
where $\bTheta$ and $\bPsi$ are sensitivity and specificity parameters necessary for modeling the imperfect measurements; The rest of parameters $\bnu=(\nu_1, \ldots, \nu_K)^\top$, $\bEta=(\eta_1, \ldots, \eta_K)^\top\in \cS_{K-1}$. Existing methods for estimating $\bpi$ in the framework of npLCM can be classified by whether or not $\bP_0$ and $\bP_{1\ell}$ assumes local independence (LI) which means measurements are independent of one another given the class ($I_i = \ell=0, 1, \ldots, L$). In Equation (\ref{eq:no_x_lkd}), LI results if and only if $\nu_1=\eta_1=1$; Otherwise, $\bnu$ and $\bEta$ account for deviations from LI given a control or disease class.

\noindent \underline{\textit{PLCM}.} $\bP_0(\bm{m})$ under the original pLCM \citep{wu2016partially} satisfies LI and equals a product of $J$ probabilities: $\bP_0(\bm{m}) = \prod_{j=1}^J\{\psi_{j}\}^{m_j}\{1-\psi_{j}\}^{1-m_j}=\Pi(\bm{m}; \bpsi)$, where $\Pi(\bm{m};\bs)=\prod_{j=1}^J \{s_{j}\}^{m_{ij}}\{1-s_{j}\}^{1-m_{ij}}$ is the probability mass function for a product Bernoulli distribution given the success probabilities $\bs=(s_1, \ldots, s_J)^\top$, $0 \leq s_j \leq 1$ and the parameters $\bpsi = (\psi_{1}, \ldots, \psi_J)^\top$ represent the positive rates absent disease, referred to as ``false positive rates" (FPRs). For example, in the PERCH data, Respiratory Syncytial Virus ({\sf RSV}) has a low observed FPR because of its rare appearance in controls' NPs; Other pathogens such as Rhinovirus ({\sf RHINO}) have higher observed FPRs. 

For $\bP_{1\ell}(\bm{m})$, the pLCM makes a key ``non-interference" assumption that disease-causing pathogen(s) are more frequently detected among cases than controls and the non-causative pathogens are observed with the same rates among cases as in controls \citep{wu2017nested}. The ``non-interference" assumption says that $\bP_{1\ell}(\bm{m})$ in a case class $\ell\neq 0$ is a product of the probabilities of measurements made ({\sf a}) on the \textit{causative} pathogen $\ell$, $\PP(M_{\ell} \mid I = \ell, Y = 1, \btheta)=\{\theta_{\ell}\}^{M_\ell}\{1-\theta_{\ell}\}^{1-M_\ell}$, where $\btheta = (\theta_1, \ldots, \theta_L)^\top$ and ({\sf b}) on the \textit{non-causative} pathogens $\PP(\bm{M}_{i[-\ell]} \mid  I_i = \ell, Y_i = 1, \bpsi_{[-\ell]})= \Pi(\bM_{[-\ell]}; \bpsi_{[-\ell]})$, where  $\ba_{[-\ell]}$ represents all but the $\ell$-th element in a vector $\ba$. The parameter $\theta_\ell$ is termed ``true positive rate" (TPR) and may be larger than the FPR $\psi_\ell$; Under the single-pathogen-cause assumption, pLCM uses $J$ TPRs $\btheta$ for $L=J$ causes and $J$ FPRs $\bpsi$.
%

\noindent \underline{\textit{NPLCM}.} To reduce estimation bias in $\bpi$ under deviations from LI, the ``nested pLCM" or npLCM extends the original pLCM to describe residual correlations among $J$ binary pathogen measurements in the controls ($I_i=0$) and in each case class ($I_i=\ell$, $\ell\neq0$) \citep{wu2017nested}. The extension is motivated by the ability of the classical LCM formulation \citep{lazarsfeld1950thelogical} to approximate any joint multivariate discrete distribution \citep{dunson2009nonparametric}. 

For $\bP_0(\bm{m})$ in the controls, the npLCM introduces $K$ subclasses; The original pLCM results if $K=1$.  Given a subclass $k$, the probability of observing $J$ binary measurements $\bM =\bm{m}$ among controls is $\bP_{0}^{(k)}(\bm{m})=\PP(\bm{M}=\bm{m} \mid Z = k, I = 0, Y = 0, \{\psi_{k}^{(j)}\}) = \Pi(\bm{m}; \bPsi_k)$, where $\bPsi_k$ is the $k$-th column of a $J$ by $K$ FPR matrix $\bPsi = \{\psi_{k}^{(j)}\}$. Since we do not observe controls' subclasses, $\bP_0$ is a weighted average of $\bP_0^{(k)}$ according to the subclass probabilities $\{\nu_k\}$: $\bP_0 = \sum_{k}^K \nu_k\bP_{0}^{(k)}$.

For $\bP_{1l}(\bm{m})$ in case class $\ell\neq 0$, the npLCM again introduces $K$ unobserved subclasses and assumes $\bP_{1\ell}$ is a weighted average of $\bP_{1\ell}^{(k)}$ according to the case subclass weights $\{\eta_k\}$: $\bP_{1\ell} = \sum_{k=1}^{K}\eta_k\bP_{1\ell}^{(k)}$. In particular, the npLCM assumes the probability of observing $\bm{M}$ in subclass $k$ in disease class $\ell \neq 0$, $\bP_{1l}^{(k)}=\PP(\bm{M} \mid Z = k, I = \ell, Y = 1)$, is a product of the probabilities of making an observation ({\sf a}) on the \textit{causative} pathogen $\ell$: $\PP(M_{\ell} \mid Y = 1,  Z = k, I = \ell, \theta_{k}^{(\ell)})=\{\theta_{k}^{(\ell)}\}^{M_\ell}\{1-\theta_{k}^{(\ell)}\}^{1-M_\ell}$ and ({\sf b}) on \textit{non-causative} pathogens $\PP(\bm{M}_{[-\ell]} \mid Y = 1, Z = k,  I = \ell, \bPsi_{k}^{([-\ell])}) =\Pi(\bM_{[-\ell]}; \bPsi_k^{([-\ell])}) =\prod_{j\neq \ell}\{\psi_{k}^{(j)}\}^{m_{j}}\{1-\psi_{k}^{(j)}\}^{1-m_{j}}$, where $\bPsi_{k}^{([-\ell])}$ is the $k$-th column of $\Psi$ excluding the $\ell$-th row. We collect the TPRs in a $J$ by $K$ TPR matrix $\bTheta = \{\theta_{k}^{(j)}\}$.  We summarize the preceding specification by $\bP_{1l}^{(k)}=\Pi(\bM;\bp_{k\ell}), \ell \neq 0$, where the vector $\bp_{k\ell} = \{p_{k\ell}^{(j)},j=1, \ldots, J\}$ represents the positive rates for $J$ measurements in subclass $k$ of disease class $\ell$:  $p_{k\ell}^{(j)}= \left\{\theta_k^{(j)}\right\}^{\ind\{j=\ell\}}\cdot\left\{\psi_k^{(j)}\right\}^{1-\ind\{j=\ell\}}$ which equals the TPR $\theta_k^{(j)}$ for a causative pathogen and the FPR $\psi_k^{(j)}$ otherwise; Here $\ind\{A\}$ is an indicator function that equals $1$ if the statement A is true and $0$ otherwise. 

The likelihood for npLCM results upon substituting $\bP_0$ and $\bP_{1\ell}$ above into Equation (\ref{eq:no_x_lkd}): $L=L_1\cdot L_0 =(\prod_{i:Y_i=1}\sum_{\ell=1}^L\pi_\ell[\cdot \sum_{k=1}^K\{\eta_k\cdot \Pi(\bM_i;\bp_{k\ell})\}])\times \prod_{i': Y_{i'}=0} \sum_{k=1}^K\nu_k\cdot \Pi(\bM_{i'}; \bPsi_k)$. Setting $\nu_1=\eta_1=1$ and $\nu_k=\eta_k=0, k\geq 2$, the special case of pLCM results. 

Similar to the pLCM, the FPRs $\bPsi$ in the npLCM are shared among controls and case classes over non-causative pathogens (via $\bp_{kl}$). Different from the pLCM, the subclass mixing weights may differ between cases ($\bEta$) and controls ($\bnu$). The special case of $\eta_k=\nu_k, k=1, \ldots, K$, means the covariation patterns among the non-causative pathogens in a disease class is no different from the controls. However, relative to controls, diseased individuals may have different strength and direction of measurement dependence in each disease class. By allowing the subclass weights to differ between the cases and the controls, npLCM is more flexible than  pLCM in referencing cases' measurements against controls. 

\section{Regression Analysis via npLCM}
\label{sec:nplcm_withX}

We extend npLCM to perform regression analysis of data $\cD = \{(\bM_i, Y_i, \bX_iY_i, \bW_i), i=1, \ldots, N\}$, where $\bX_i=(X_{i1}, \ldots, X_{ip})^\top$ are covariates that may influence case $i$'s etiologic fractions and $\bW_i = (W_{i1}, \ldots, W_{iq})^\top$ is a possibly different vector of covariates that may influence the subclass weights among the controls and the cases; Let the continuous covariates comprise the first $p_1$ and $q_1$ elements of $\bX_i$ and $\bW_i$, respectively. A subset of $\bX_i$ may be available from the cases only. We let $\bX_iY_i=\bm{0}_{p\times 1}$ if $Y_i=0$ so that all the covariates for a control subject are included in $\bW_i$; Let $\bX_iY_i=\bX_i$ for a case subject.  For example, healthy controls have no disease severity information. We let three sets of parameters in an npLCM (\ref{eq:no_x_lkd}) depend on the observed covariates: ({\sf a}) the etiology regression function among cases, $\{\pi_\ell(\bx), \ell\neq 0\}$, which is of primary scientific interest, ({\sf b}) the conditional probability of measurements $\bm{m}$ given covariates $\bw$ in case classes: $\bP_{1\ell}(\bm{m}; \bw)=[\bM=\bm{m} \mid \bW=\bw,I=\ell]$, $\ell = 1, \ldots, L$, ({\sf c}) and in the controls $\bP_0(\bm{m}; \bw)=[\bM=\bm{m} \mid \bW=\bw,I=0]$;  We keep the specifications for the TPRs and FPRs ($\bTheta$, $\bPsi$) as in the original npLCM.

\subsection{Disease Etiology Regression}

$\pi_\ell(\bX)$ is the primary target of inference. Recall that $I_i=\ell$ represents case $i$'s disease being caused by pathogen $\ell$. We assume this event occurs with probability $\pi_{i\ell}$ that depends upon covariates. In our model, we use a multinomial logistic regression model $\pi_{i\ell} = \pi_\ell(\bX_i)= \exp\{\phi_\ell(\bX_i)\}/\sum_{\ell'=1}^L\exp\{\phi_{\ell'}(\bX_i)\}$, $\ell =1, ..., L$, where  $\phi_{\ell}(\bX_i)-\phi_L(\bX_i)$ is the log odds of case $i$ in disease class $\ell$ relative to $L$: $\log {\pi_{i\ell}}/{\pi_{iL}}$. Without specifying a baseline category, we treat all the disease classes symmetrically which simplifies prior specification. We further assume additive models for \(\phi_\ell(\bm{x}; \bGamma_\ell^\pi) = \sum_{j=1}^{p_1} f^\pi_{\ell j}(x_j; \bbeta^\pi_{\ell j})+\tilde{\bx}^\top\bgamma^\pi_\ell\), where $\tilde{\bx}$ is the subvector of the predictors $\bx$ that enters the model for all disease classes as linear predictors and $\bGamma_\ell^\pi = (\bbeta^\pi_{\ell j}, \bgamma_\ell^\pi)$ collects all the parameters. For covariates such as enrollment date that serves as a proxy for factors driven by seasonality, nonlinear functional dependence is expected. We use B-spline basis expansion to approximate $f_{\ell j}^\pi(\cdot)$ and use P-spline for estimating smooth functions \citep{Lang2004}. Finally, we specify the distribution of case measurements $\bM$ given disease class $I$, covariates $\bX$ and $\bW$. We extend the case likelihood $L_1$ in an npLCM (\ref{eq:no_x_lkd}) to let the subclass weights depend on covariates $\bW$: \(
P(\bM \mid \bW, I=\ell, Y=1)  = \sum_{k=1}^K \eta_{k}(\bW)\cdot \Pi\left(\bM; \bp_{k\ell }\right), \ell=1, \ldots, L\). Integrating over $L$ unobserved disease classes, we obtain the likelihood function for the cases that incorporates covariates $\{\bX_i, \bW_i\}$:
\begin{align}
L^{\tiny \sf reg}_1 & = \prod_{i:Y_i=1} \left\{\sum_{\ell=1}^L \left[\pi_\ell(\bX_i; \bGamma^\pi_\ell) \sum_{k=1}^K\left\{\eta_{ik}\cdot\Pi(\bM_i; \bp_{k\ell})\right\}\right]\right\},\label{eq:case_likelihood_reg}
\end{align}
where $\eta_{ik}=h_k(\bW_i; \bGamma^\eta_k)$ and $\bGamma_k^\eta$ are the regression parameters; The form of $h_k$ is introduced in the model for controls.

\subsection{Covariate-dependent reference distribution}

Data from controls provide requisite information about the specificities and covariations at distinct covariate values, necessitating adjustment in an npLCM analysis. For example, factors such as enrollment date is a proxy for season and may influence the background colonization rates and interactions of some pathogens that circulate more during winter \citep{obando2018respiratory, nair2011global}. We propose a novel approach to estimating the reference distribution of measurements that may depend on covariates using control data.

The regression model for a control subject is a mixture model with covariate-dependent mixing weights $\nu_{k}(\bW)$: \(
\PP(\bm{M}\mid \bW, Y=0) = \sum_{k=1}^K\nu_{k}(\bW)\Pi(\bm{M}; \bPsi_k),\)
where FPRs $\bPsi_k = (\psi^{(1)}_k, \ldots, \psi^{(J)}_k)^\top$ do not depend on covariates and the vector $\bnu(\bW)=\left(\nu_{1}(\bW), \ldots, \nu_K(\bW)\right)^\top$ lies in a $(K-1)$-simplex $\cS_{K-1}$. We discuss the FPRs $\{\bPsi_k\}$ and the subclass weight functions $\{\nu_k(\bW)\}$ in order. 



Firstly, constant FPR profiles $\{\bPsi_k\}$ enable coherent interpretation across individuals with different covariate values \citep{erosheva2007describing}. FPR profile $k$ receives a weight of $\nu_k(\bW_i)$ for a control subject $i$  with covariates $\bW_i$. The \textit{marginal} FPRs in the controls $\PP(\bM_{j} = 1 \mid  \bW, Y = 0, \bPsi) = \sum_{k=1}^K\nu_k(\bW)\psi_{k}^{(j)} \in [\min_k \psi_{k}^{(j)}, \max_k \psi_{k}^{(j)}]$, $j=1, \ldots, J$, also  depend on $\bW$. Consequently, observed marginal control positive curve for a pathogen informs how different the FPRs $\bPsi_k^{(j)}$ are across the subclasses. For example, if the NPPCR measure of pathogen {\sf A} shows strong seasonal trends among the controls, the estimated FPRs will be more variable across the subclasses. And the subclass with a high FPR will receive a larger weight during seasons with higher carriage rates in controls. The control model reduces to special cases, with covariate-independent $\nu_k(\bW)\equiv \nu_k$, $k=1, \ldots, K$, resulting in the $\bP_0$ in a $K$-subclass npLCM without covariates; A further single-subclass constraint ($K=1$) gives the $\bP_0$ in the original pLCM.

Secondly, we parameterize the case and control  subclass weight regressions $\eta_k(\bW)$ and $\nu_k(\bW)$ using the same regression form $h_k(\bW; \cdot)$ but different parameters.

\noindent \underline{\textit{Control subclass weight regression}.} We rewrite the subclass weights $\nu_k(\cdot), k=1, \ldots, K$, using a stick-breaking parameterization. Let $g(\cdot): \RR \mapsto [0,1]$ be a link function. Let $\alpha_{ik}$ be subject $i$'s linear predictor at stick-breaking step $k=1, \ldots, K-1$. Using the stick-breaking analogy, we begin with a unit-length stick, break a segment of length $g(\alpha^\nu_{i1})$ and continue breaking a fraction $g(\alpha^\nu_{i2})$ from the remaining $\{1-g(\alpha^\nu_{i1})\}$ and so on;  At step $k$, we break a fraction $g(\alpha^\nu_{ik})$ from what is left in the preceding $k-1$ steps resulting in the $k$-th stick segment $k$ of length $\eta_{ik}=g(\alpha^\nu_{ik})\prod_{s<k} \{1-g(\alpha^\nu_{is})\}$; We stop until $K$ sticks of variable lengths result. In this paper, we use the logistic function $g(\alpha) = 1/\left\{1+\exp(-\alpha)\right\}$ which is consistent with the multinomial logit regression for $\pi_\ell(\cdot)$ so that the priors of the coefficients $\bGamma_k^\nu$ and $\bGamma_\ell^\pi$ can be similar (Supplementary Materials \ref{sec:prior.smooth}). Generalization to other link functions such as the probit function is straightforward \citep[e.g.,][]{rodriguez2011nonparametric}. We use this parameterization to introduce a novel shrinkage prior on a simplex for the subclass weights $\{\nu_k(\bW)\}$ (see Supplementary Materials \ref{sec:prior.few.subclasses}) which encourages fewer than $K$ effective subclasses, or ``$m$-sparse" shrinkage prior on the simplex. This provides parsimonious approximation to the conditional distribution of control measurements $\PP(\bM \mid \bW, Y = 0, \{\nu_k(\cdot)\},\bPsi)$ using a few subclasses. 

In our analysis, we use generalized additive models \citep{hastie1986generalized} for the $k$-th linear predictor \(
\alpha^\nu_{ik} = \alpha^\nu_{k}(\bW_i = \bm{w};\bGamma^\nu_k) = \mu_{k0}+\sum_{j=1}^{q_1}f_{kj}(w_j; \bbeta^\nu_{kj})+\tilde{\bw}^\top\bgamma^\nu_k,
\) for $k=1, \ldots, K-1$. We have parameterized the possibly nonlinear $f_{kj}(\cdot)$ using B-spline basis expansions with coefficients $\bbeta^\nu_{kj}$; $\tilde{\bw}^\top\bgamma^\nu_k$ are the linear effects of a subset of predictors which can include an intercept and $\tilde{\bw}$ is a subvector of predictors $\bw$; Let $ \bGamma^\nu_k = \{\mu_{k0}, \{\bbeta^\nu_{kj}\}, \bgamma_k^\nu\}$ collect all the regression parameters. Following \citet{Lang2004}, we constrain $\{f_{kj}, j=1, \ldots, J\}$ to have zero means for statistical identifiability. Supplementary Materials \ref{sec:prior.smooth} provides the technical details about the parameterization of $f_{kj}$. 

The subclass-specific intercepts $\{\mu_{k0}\}$ globally control the magnitudes of the linear predictors. We hence propose priors on $\{\mu_{k0}\}$ to \textit{a priori} encourage few subclasses (see Supplementary Materials \ref{sec:prior.few.subclasses}). In particular, a large positive intercept $\mu_{k0}$ makes $g(\alpha^\nu_{ik})\approx 1$ and hence breaks nearly the entire remaining stick after the $(k-1)$-th stick-breaking. Since the stick-breaking parameterization one-to-one maps to a classical latent class regression model formulation for the control data, the linear predictor $\alpha^\nu_{ik}$ and the sum $\mu_{k0}+\gamma^\nu_{k0}$ are identifiable except in a Lebesgue zero set of parameter values, or ``generic identifiability" \citep{huang2004building}. Consequently, even if the intercept $\mu_{k0}$ is not statistically identified if $\tilde{\bw}$ includes an intercept $\gamma^\nu_{k0}$, the MCMC samples of the statistically identifiable functions can provide valid posterior inferences \citep{Carlin2009}. We write the control likelihood with covariates $\bW_i$ as $L^{\tiny \sf reg}_0 = \prod_{i:Y_{i}=0} \sum_{k=1}^Kh_k(\bW_{i};\bGamma_k^\nu)\Pi(\bM_{i}; \bPsi_k)$. Supplementary Materials \ref{sec:remark} provides further remarks on the assumption for introducing covariates into the control model.




\noindent \underline{\textit{Case subclass weight regression}.} The subclass weight regression functions for cases $\{\eta_k(\bW)\}$ are also specified via a logistic stick-breaking regression as in the controls but with different parameters: $\eta_{ik} = g(\alpha^\eta_{ik}) \prod_{s<k} \{1-g(\alpha_{is}^\eta)\}$, $k=1, \ldots, K-1$. Since given the TPRs and the FPRs, the subclass weights fully determine the joint distribution $[\bM\mid \bW, I=\ell\neq0]$ hence the measurement dependence in each class, we let  $\eta_k(\bw)$ and $\nu_k(\bw)$ be different between cases and controls for any $\bw$. 


Let the $k$-th linear predictor $\alpha_{ik}^\eta = \alpha_{k}^\eta(\bW_i=\bw; \bGamma_k^\eta) =  \mu_{k0}+\sum_{j=1}^{q_1} f_{kj}(w_j; \bbeta_{kj}^\eta)+\tilde{\bw}^\top\bgamma_k^\eta$, where $\bGamma^\eta_k = \{\mu_{k0},\{\bbeta_{kj}^\eta\}, \bgamma_k^\eta\}$ are the regression parameters that differ from the control counterpart ($\bGamma_k^\nu$). In particular, we approximate $f_{kj}(\cdot), j=1, \ldots, J$, here using the same set of B-spline basis functions as in the controls but estimate a different set of basis coefficients $\bbeta_{kj}^\eta$. In addition, we have directly used the intercepts $\{\mu_{k0}\}$ from the control model to ensure only important subclasses in the controls are used in the cases. For example, absent covariates $\bW$, a large and positive $\mu_{k0}$ effectively halts the stick breaking procedure at step $k$ for the controls ($\nu_{k+1}\approx 0$); Applying the same intercept $\mu_{k0}$ to the cases makes $\eta_{k+1}\approx 0$.

Combining the case ($L^{\tiny \sf reg}_1$) and control likelihood ($L^{\tiny \sf reg}_0$) with covariates, we obtain the joint likelihood for the regression model $L^{\tiny \sf reg} = L^{\tiny \sf reg}_1 \times L^{\tiny \sf reg}_0$. 

\begin{remark}
\label{remark:marginal}
Under an assumption {\sf (A1)}: the case subclass weights are constant over covariates: $\eta_k(\cdot) \equiv \eta_k$, $k=1, \ldots, K$, the regression model reduces to an npLCM model without covariates upon integration over a distribution of covariates $\bX$. To see this, the case and control likelihood functions $L^{\tiny \sf reg}_1$ and $L^{\tiny \sf reg}_0$ integrate to \(L_1^*= \prod_{i:Y_i=1} \sum_{\ell=1}^L \pi_\ell^\ast \sum_{k=1}^K\eta_k\Pi(\bM_i; \bp_{k\ell}),\) and \(L_0^*=\prod_{i:Y_i=0}\sum_{k=1}^K\nu^*_k \Pi(\bM_i; \bPsi_k)\), respectively; Here $\pi_\ell^\ast=\int \pi_{\ell}(\bX) \mathrm{d} G(\bX)$ and $\nu^\ast_k = \int \nu_k(\bW) \mathrm{d}H(\bW)$ where $G$ and $H$ are probability or empirical distributions of $\bX$ and $\bW$, respectively. The mathematical equivalence enables valid inference about the overall PEFs $\bpi^\ast$ omitting $\bX$ and $\bW$ (see Supplementary Materials \ref{sec:valid_though_nox} for an example). The no-covariate analysis becomes deficient under deviations from {\sf (A1)}; Section \ref{sec:simulation} provides examples.
\end{remark}

\subsubsection{Priors and Posterior Inference}
The unknown parameters include the coefficients in the etiology regression ($\{\bGamma_\ell^\pi\})$, the subclass mixing weight regression for the cases ($\{\bGamma_k^\eta\}$) and the controls  ($\{\bGamma^\nu_k\}$), the true and false positive rates $(\bTheta=\{\theta_{k}^{(j)}\}, \bPsi=\{\psi_{k}^{(j)}\})$. With typical samples sizes about $500$ controls and $500$ cases in each study site, the number of parameters in controls likelihood $L_0$ ($>JKCp$) easily exceeds the number of distinct binary measurement patterns observed. To overcome potential overfitting and increase model interpretability, we \textit{a priori} place substantial probabilities on models with the following two features: ({\sf a}) Few non-trivial subclasses via a novel additive half-Cauchy prior for the intercepts $\{\mu_{k0}\}$, and ({\sf b}) for a continuous variable, smooth regression curves $\pi_\ell(\cdot)$, $\nu_k(\cdot)$ and $\eta_k(\cdot)$ by Bayesian Penalized-splines (P-splines) \citep{Lang2004} combined with shrinkage priors on the spline coefficients  \citep{ni2015bayesian} to encourage towards constant values, $\eta_k(\cdot) = \eta_k, \nu_k(\cdot)=\nu_k, k=1, \ldots, K$, which reduces to the original npLCM. Supplementary Materials \ref{sec:prior_distribution} details the prior specifications.

We  use the Markov chain Monte Carlo (MCMC) algorithm to draw samples of the unknowns to approximate their joint posterior distribution \citep{Gelfand1990}. Flexible posterior inferences about any functions of the model parameters and individual latent variables are available by plugging in the posterior samples of the unknowns. For example, the posterior samples of the case positive rate curve for pathogen $j$ help evaluate model fit. The red bands in Row 1 of  Figure \ref{fig:seasonality_simulation} are posterior $95\%$ credible bands obtained by substituting relevant parameters with their sampled values across MCMC iterations in $\PP(M_{\ell} = 1 \mid \bx,\bw, Y=1) = \pi_{\ell}(\bw; \bGamma_\ell^\pi) \sum_{k=1}^K h_k(\bw;\bGamma_k^\eta)\theta_k^{(\ell)}+\{1-\pi_{\ell}(\bx; \bGamma_\ell^\pi)\} \sum_{k=1}^K h_k(\bw;\bGamma_k^\eta)\psi_k^{(\ell)}$. The npLCMs with or without covariates are fitted using a free and publicly available  R package \verb"baker" (\url{https://github.com/zhenkewu/baker}). \verb"Baker" calls an external automatic Bayesian model fitting software \verb"JAGS 4.2.0" \citep{plummer2003jags} from within R and provides functions to visualize the posterior distributions of the unknowns (e.g., the PEFs and cases' latent disease class indicators) and perform posterior predictive model checking \citep{gelman1996posterior}. Supplementary Materials \ref{sec:mcmc} details the convergence diagnostics.


\section{Simulations}
\label{sec:simulation}
We simulate case-control bronze-standard (BrS) measurements along with observed continuous and/or discrete covariates under multiple combinations of true model parameter values and sample sizes that mimic the motivating PERCH study. In {\sf Simulation I}, we illustrate flexible statistical inferences about the PEF functions $\{\pi_\ell(\cdot)\}$. In {\sf Simulation II}, we focus on the overall PEFs that quantify the overall cause-specific disease burdens in a population which are of policy interest. Let $\pi_\ell^\ast$ be an empirical average of $\pi_\ell(\bX)$, $\ell=1, \ldots, L$. We compare the frequentist properties of the posterior mean $\bpi^\ast$ obtained from analyses with or without covariate \citep{little2011calibrated}. Regression analyses reduce estimation bias, retain efficiency and provide more valid frequentist coverage of the $95\%$ CrIs. The relative advantage varies by the true data generating mechanism and sample sizes. 

In all analyses here, we use a working number of $K^\ast$ subclasses, with independent {\sf Beta}(7.13,1.32) TPR prior distributions that match 0.55 and 0.99 with the lower and upper $2.5 \%$ quantiles, respectively; We specify {\sf Beta}(1,1) for the identifiable FPRs. The priors for the regression coefficients follow the specifications in Supplementary Materials \ref{sec:prior_distribution}.


\noindent \underline{{\it Simulation I}.}  We demonstrate that the inferential algorithm recovers the true PEF functions $\{\pi^0_\ell(\bX)\}$. We simulate $N_d=500$ cases and $N_u=500$ controls for each of two levels of $S$ (a discrete covariate) and uniformly sample the subjects' enrollment dates over a period of $300$ days. Supplementary Materials \ref{sec:simulation_main} specifies the true data generating mechanism and the regression specifications. Based on the simulated data, pathogen {\sf A} has a bimodal positive rate curve mimicking the trends observed of {\sf RSV} in one PERCH site; other pathogens have overall increasing positive rate curves over enrollment dates. We set the simulation parameters in a way that the \textit{marginal} control rate may be higher than cases for small $t$'s (impossible under the more restrictive pLCM). Row 2 of Figure \ref{fig:seasonality_simulation} visualizes for the $9$ causes (by column), the posterior means (thin black line) and $95\%$ CrIs (gray bands) for the etiology regression curves ${\pi}_\ell(\cdot)$ are close to the simulation truths $\pi^0_\ell(\cdot)$. Supplementary Materials \ref{sec:extra_simulation} provides additional simulation results to assess the recovery of the true $\pi^0_\ell(X)$ for a discrete covariate $X$.



\begin{figure}[!h]
\centering
\includegraphics[width=\textwidth]{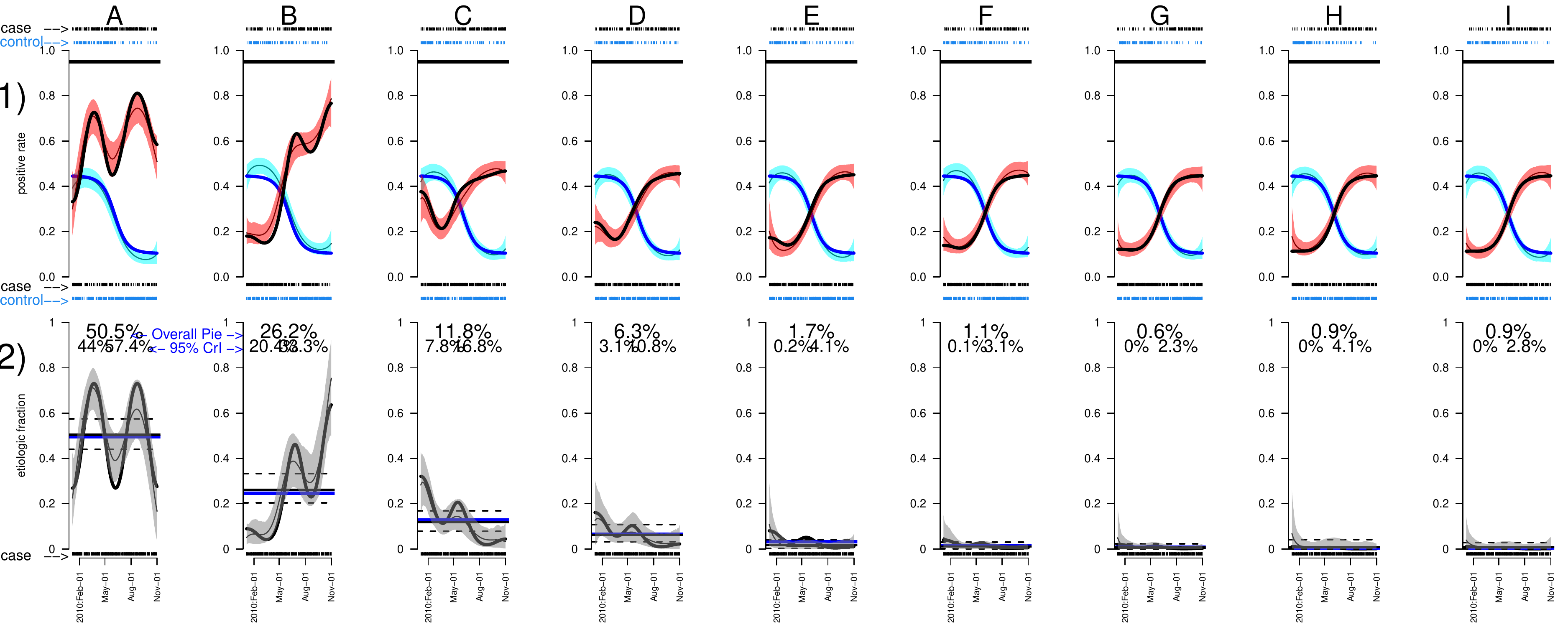}
\caption{Row 2) For each of the $9$ causes (by column) in Simulation {\sf I}, the posterior mean (thin black curves) and pointwise $95\%$ credible bands (gray bands) for the etiology regression curves ${\pi}_\ell(x)$ are close to the simulation truths $\pi^0_\ell(x)$. In row 1), the fitted case (red) and control (blue) positive rate curves are shown with the posterior mean curves (solid black curves) and pointwise $95\%$ credible bands (shaded); The rug plots show the positive (top) and negative (bottom) measurements made on cases and controls on the enrollment dates. The solid horizontal lines in row 1 indicate the true TPRs.}
\label{fig:seasonality_simulation}
\end{figure}

\noindent \underline{{\it Simulation II}.} We show the regression model accounts for population stratification by covariates hence reduces the bias of the posterior mean $\{\hat{\pi}_\ell^\ast\}$ in estimating the overall PEFs ($\bpi^\ast$) and produces more valid $95\%$ CrIs. We illustrate the advantage of the regression approach under simple scenarios with a single two-level covariate $X\in\{1,2\}$; We let $W=X$. We perform npLCM regression analysis with $K^*=3$ for each of $R=200$ replication data sets simulated under each of $48$ scenarios detailed in Supplementary Materials \ref{sec:simulation_main} that correspond to distinct numbers of causes, sample sizes, relative sizes of PEF functions (rare versus popular etiologies), signal strengths (more discrepant TPRs and FPRs indicate stronger signals, \citet{wu2016partially}), and effects of $W$ on $\{\nu_k(W)\}$ and $\{\eta_k(W)\}$.

In estimating $\pi^*_\ell$, we evaluate the bias $\widehat{\pi^\ast_\ell}-\pi_\ell^{0\ast}$, where $\pi_\ell^{0\ast}=N_1^{-1}\sum_{i: Y_i=1}\pi^0_\ell(\bX_i)$ is the true overall PEF, and $\widehat{\pi^\ast_\ell} = N_1^{-1}\sum_{i: Y_i=1} \widehat{\pi}_\ell(\bX_i)$ is an empirical average of the posterior mean PEFs at $\bX_i$. We also evaluate the empirical coverage rates of the $95\%$ CrIs.

The regression model incorporates covariates and performs better in estimating $\bpi^\ast$ than a model omitting covariates. For example, Figure \ref{fig:res_relbias} shows for $J=6$ that, relative to no-covariate npLCM analyses, regression analyses produce posterior means that on average have negligible relative biases (percent difference between the posterior mean and the truth relative to the truth) for each pathogen across simulation scenarios. As expected, we observe slight relative biases from the regression model in the bottom two rows of Figure \ref{fig:res_relbias}, because the informative TPR prior {\sf Beta}(7.13,1.32) has a mean value lower than the true TPR $0.95$; A more informative prior further reduces the relative bias; See additional simulations in Supplementary Materials \ref{sec:extra_simulation} on the role of informative TPR priors. Figure \ref{fig:res_coverage} regression analyses also produce $95\%$ CrIs for $\pi^\ast_\ell$ that have more valid empirical coverage rates in all scenarios. Misspecified models without covariates concentrate the posterior distribution away from the true overall PEFs, resulting in large biases that dominate the posterior uncertainty of $\pi_\ell^\ast$ which is evident from the more severe undercoverages with higher TPRs and lower FPRs (row 3 and 4 versus row 1 and 2, Figure \ref{fig:simulation2res}). 

\begin{figure}[!htp]
\captionsetup{width=\linewidth}
\centering
\addtocounter{figure}{0} 
\raisebox{-.5in}{
\subfigure[Percent relative bias]{
 \includegraphics[width=0.8\linewidth]
{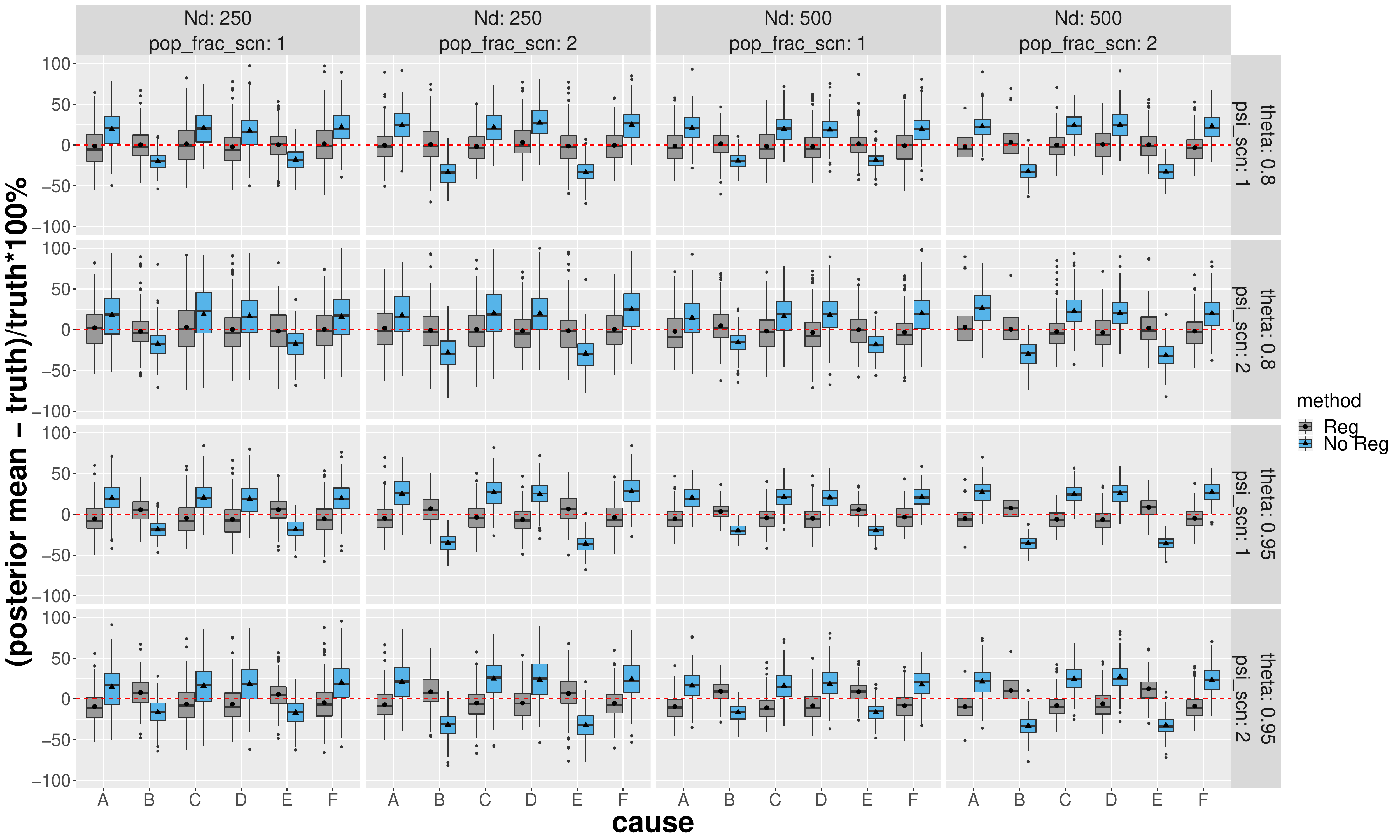}
\label{fig:res_relbias}
}
}
\hspace*{.2in}
{\subfigure[Empirical coverage rates]{
\includegraphics[width=0.8\linewidth]{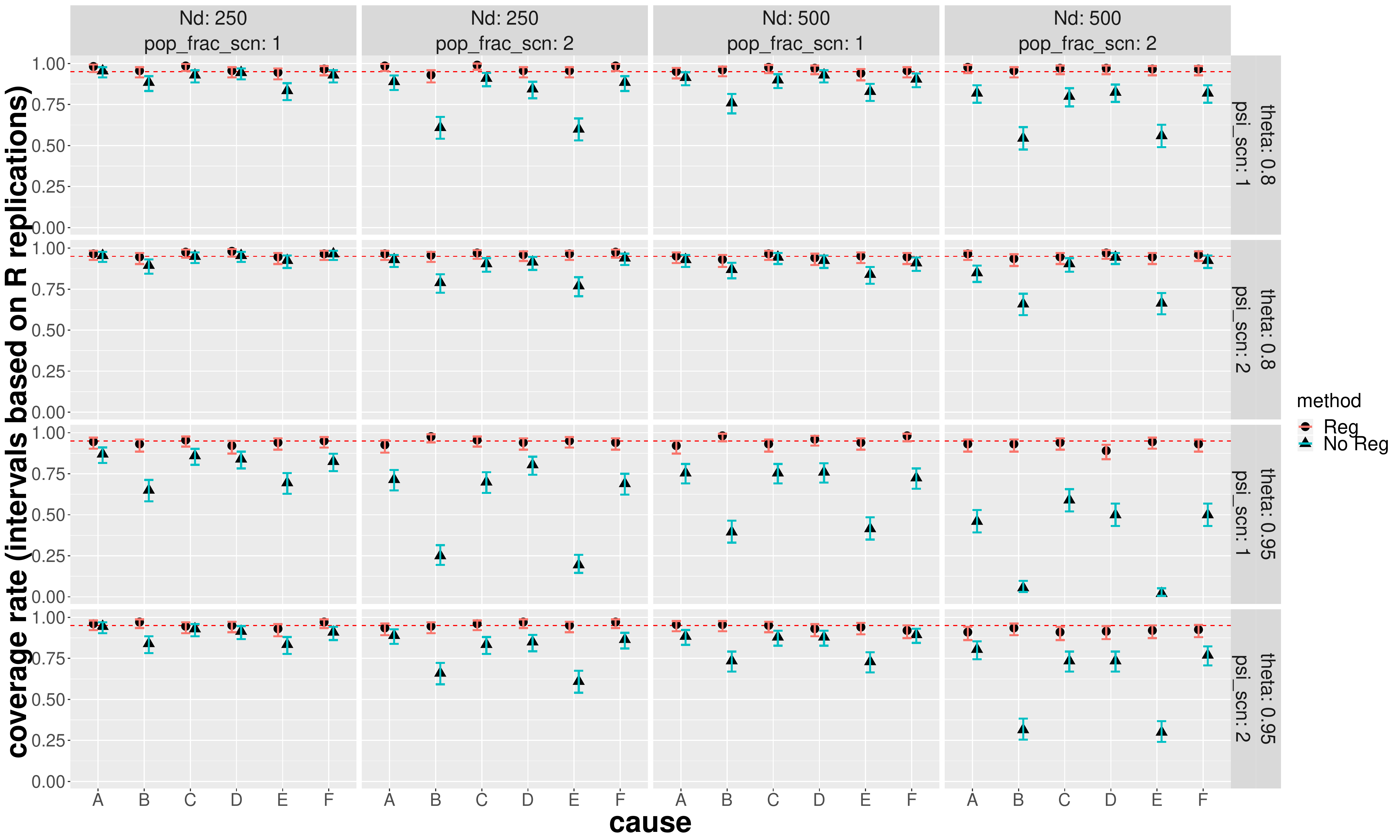}
\label{fig:res_coverage}
}
}
\addtocounter{figure}{0} 
\caption{The regression analyses produce less biased posterior mean estimates and more valid empirical coverage rates for $\pi_\ell^\ast$ over $R=200$ replications in {\sf Simulation II} with $J=6$.  Each panel corresponds to one of $16$ combinations of true parameter values and sample sizes. \textit{Top}) Each boxplot (left: regression; right: no regression) shows the distribution of the percent relative bias of the posterior mean in estimating the overall PEF $\pi^\ast_\ell$ for six causes ({\sf A} - {\sf F}); The red horizontal dashed lines indicate zero bias. \textit{Bottom}) Each dot or triangle indicates the empirical coverage rate of the $95\%$ CrIs produced by analyses with regression ($\bullet$) or without regression ($\blacktriangle$); The nominal $95\%$ rate is marked by horizontal red dashed lines. Since each coverage rate for $\pi_\ell^\ast$ is computed from $R=200$ binary observations, the truth being covered or not, a $95\%$ CI is also shown.}
\label{fig:simulation2res}
\end{figure}


\vspace{-.20in}
\section{Regression Analysis of PERCH Data}
\label{sec:perch}

We restrict attention in this regression analysis to $494$ cases and $944$  controls from one of the PERCH study sites in the Southern Hemisphere that collected information on enrollment date ($t$, August 2011 to September 2013; standardized), age (dichotomized to younger or older than one year), disease severity for cases (severe or very severe), HIV status (positive or negative) and presence or absence of seven species of pathogens (five viruses and two bacteria, representing a subset of pathogens evaluated) in nasopharyngeal (NP) specimens tested with polymerase chain reaction (PCR), or NPPCR (bronze-standard, BrS); We also include in the analysis the blood culture (BCX, silver-standard, SS) results for two bacteria from cases only. Detailed analyses of the entire data are reported in \citet{o2019aetiology}. 

Table \ref{table:x_freq} shows the observed case and control frequencies by age, disease severity and HIV status. The two strata with the most subjects are severe pneumonia children who were HIV negative and under or above one year of age. Some low or zero cell counts preclude fitting npLCMs by stratum. Regression models with additive assumptions among the covariates can borrow information across strata and stabilize the PEF estimates. Supplemental Figure S5 shows summary statistics for the NPPCR (BrS) and BCX (SS) data including the positive rates in the cases and the controls and the conditional odds ratio (COR) contrasting the case and control rates adjusting for the presence or absence of other pathogens (NPPCR only). 

For NPPCR, pathogens {\sf RSV} and \textit{Haemophilus influenzae} ({\sf HINF}) are detected with the highest positive rates among cases: $29.3\%$ and $34.1\%$, respectively, which are higher than the corresponding control rates ($3.1\%$ and $21.7\%$). The CORs are  large, $14~(95\% \text{CI:}~9.4,21.6)$ for {\sf RSV} and $1.8~(95\% \text{CI:}~1.3,2.3)$ for {\sf HINF}, indicating etiologic importance. Adenovirus ({\sf ADENO}) also has a statistically significant COR of $1.5~(95\% \text{CI:}~1.1,2.2)$. Human metapneumovirus type A or B ({\sf HMPV\_A\_B}) and Parainfluenza type 1 virus ({\sf PARA\_1}) have larger positive and statistically significant CORs of $2.6 ~(95\% \text{CI:}~ 1.5,4.4)$ and $6.4~(95\% \text{CI:}~2.3,20.3)$. However, the two pathogens rarely appear in cases' nasal cavities ({\sf HMPV\_A\_B}: $6.8\%$, {\sf PARA\_1}: $2.3\%$), which in light of high sensitivities $(50\sim90)\%$ means non-primary etiologic roles. For the rest of pathogens, we observed similar case and control positive rates as shown by the statistically non-significant CORs  ({\sf RHINO} (case: $21.4\%$; control: $19.9\%$) and \textit{Streptococcus pneumoniae} ({\sf PNEU}) (case: $14.4\%$; control: $9.9\%$). Similar to \citet{wu2017nested}, we integrate case-only SS measurements for {\sf HINF} and {\sf PNEU} by using informative priors of the sensitivities (e.g., from vaccine probe studies e.g., \citet{feikin2014use}) to adjust the PEF estimates in a coherent Bayesian framework. It is expected that the rare detection of the two bacteria, $0.4\%$ for {\sf HINF} and $0.2\%$ for {\sf PNEU} from SS data, will lower their PEF estimates relative to the ones obtained from an NPPCR-only analysis.

We include in the regression analysis a cause ``{\sf Not Specified (NoS)}" to account for true pathogen causes other than the seven pathogens. We incorporate the prior knowledge about the TPRs of the NPPCR measures from laboratory experts. We set the Beta priors for sensitivities by $a_{\theta} = 126.8$ and $b_{\theta}=48.3$, so that the $2.5\%$ and $97.5\%$ quantiles match the lower and upper ranges of plausible sensitivity values of $0.5$ and $0.9$, respectively. We specify the {\sf Beta}(7.59,58.97) prior for the two TPRs of SS measurements similarly but with a lower range of $5-20\%$. We use a working number of subclasses $K=5$. In the etiology regression model $f^\pi_{\ell j}(t)$, we use 7 d.f. for B-spline expansion of the additive function for the standardized enrollment date $t$ at uniform knots along with three binary indicators for age older than one, very severe pneumonia, HIV positive; In the subclass weight regression model $h_k(\bW; \cdot)$, we use 5 d.f. for the standardized enrollment date $t$ with uniform knots and two indicators for age older than one and HIV positive. The prior distributions for the etiology and subclass weight regression parameters follow the specification in Supplementary Materials \ref{sec:prior_distribution}.

\begin{table}[H]
\centering
\captionsetup{width=0.95\linewidth}
\caption{The observed count (frequency) of cases and controls by age, disease severity and HIV status (1: yes; 0: no). The marginal fractions among cases and controls for each covariate are shown at the bottom. Results from the regression analyses are shown in Figure \ref{fig:data_results} for the first two strata.}

\scalebox{0.9}{
\begin{tabular}{cccrr}
\hline
\multicolumn{1}{l}{age $\geq 1$} & \multicolumn{1}{l}{very severe (VS)} & \multicolumn{1}{l}{HIV positive} & \multicolumn{1}{l}{$\#$ cases ($\%$)} & \multicolumn{1}{l}{$\#$ controls ($\%$)} \\ 
\multicolumn{1}{l}{}             & \multicolumn{1}{c}{(case-only)}                 & \multicolumn{1}{r}{}             & total: 524 (100)                         & total: 964 (100)                            \\\hline\hline
\rowcolor[HTML]{C0C0C0} 
0                                & 0                                    & 0                                & 208 (39.7)                            & 545 (56.5)                               \\
\rowcolor[HTML]{C0C0C0} 
1                                & 0                                    & 0                                & 72 (13.7)                             & 278 (28.8)                               \\
0                                & 1                                    & 0                                & 116 (22.1)                            & 0                                        \\
1                                & 1                                    & 0                                & 33 (6.3)                              & 0                                        \\
0                                & 0                                    & 1                                & 37 (7.1)                              & 85 (8.8)                                 \\
1                                & 0                                    & 1                                & 24 (4.5)                              & 51 (5.3)                                 \\
0                                & 1                                    & 1                                & 25 (4.8)                              & 0                                        \\
1                                & 1                                    & 1                                & 3 (0.6)                               & 0\\\hline 
\multicolumn{1}{r}{case: $25.2\%$ }                               & $34.5\%$                                    & $17.0\%$                                &                                & \\  
\multicolumn{1}{r}{control: $34.3\%$}                                & -                                    & $14.1\%$                                &                              &  \\\hline 
\end{tabular}
}

\label{table:x_freq}
\end{table}

The regression analysis produces seasonal estimates of the PEF function for each cause that varies in trend and magnitude among the eight strata defined by age, disease severity and HIV status. Figure \ref{fig:data_results} shows among two age-HIV-severity strata the posterior mean curve and $95\%$ pointwise credible bands  of the etiology regression functions $\pi_{\ell}(t, {\sf age}, {\sf severity}, {\sf HIV})$ as a function of $t$. For example, among the younger, HIV negative and severe pneumonia children (Figure \ref{fig:younger}), the PEF curve of {\sf RSV} is estimated to have a prominent bimodal temporal pattern that peaked at two consecutive winters in the Southern Hemisphere (June 2012 and 2013). Other single-pathogen causes {\sf HINF}, {\sf PNEU}, {\sf ADENO}, {\sf HMPV\_A\_B} and {\sf PARA\_1} have overall low and stable PEF curves across seasons. The estimated PEF curve of {\sf NoS} shows a trend with a higher level of uncertainty that is complementary to {\sf RSV} because given any enrollment date the PEFs of all the causes sum to one. In contrast, Figure \ref{fig:older} shows a lower degree of seasonal variation of {\sf RSV} PEF curve among the older, HIV negative and severe pneumonia children.

\begin{figure}[h]
\captionsetup{width=0.95\linewidth}
\centering
\addtocounter{figure}{0} 
\raisebox{-.5in}{
\subfigure[ \textbf{Age $\leq 1$ year}, severe pneumonia, HIV negative]{
 \includegraphics[width=\linewidth]
{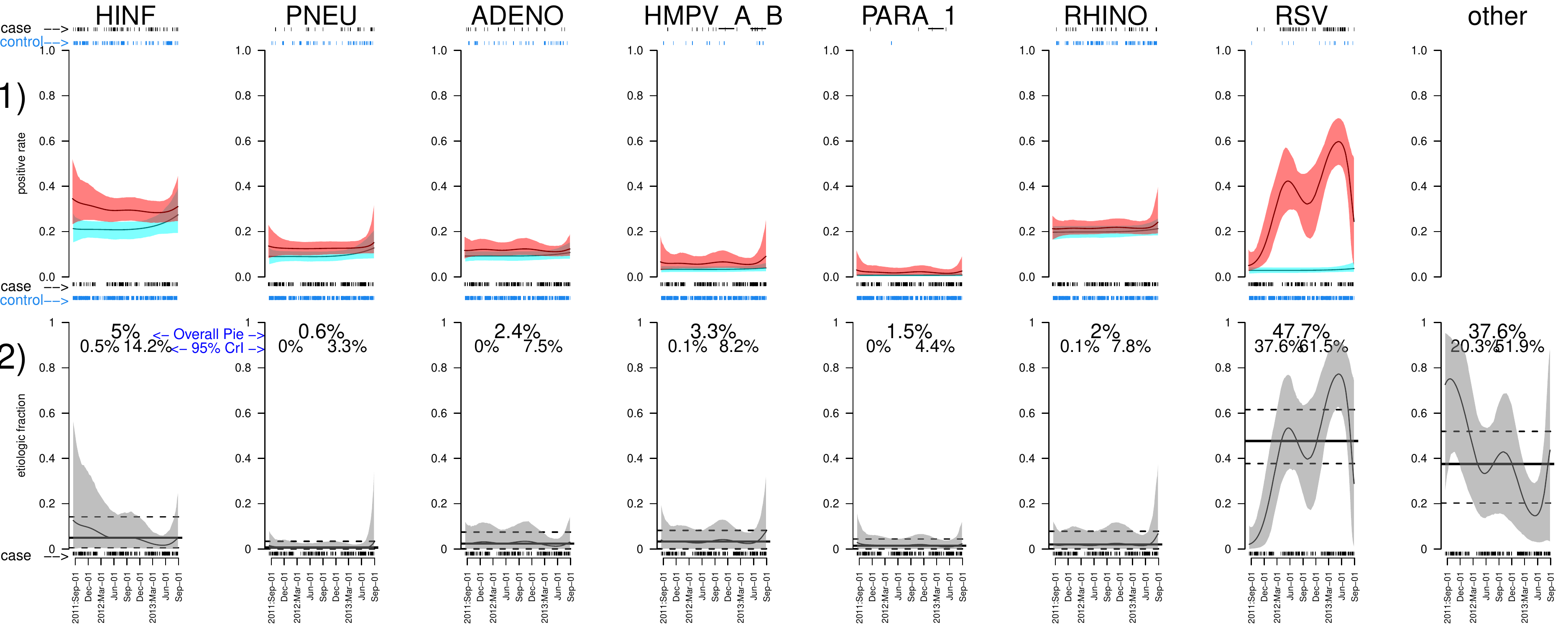}
\label{fig:younger}
}
}
\hspace*{.2in}
{\subfigure[\textbf{Age $> 1$ year}, severe pneumonia, HIV negative]{
\includegraphics[width=0.95\linewidth]{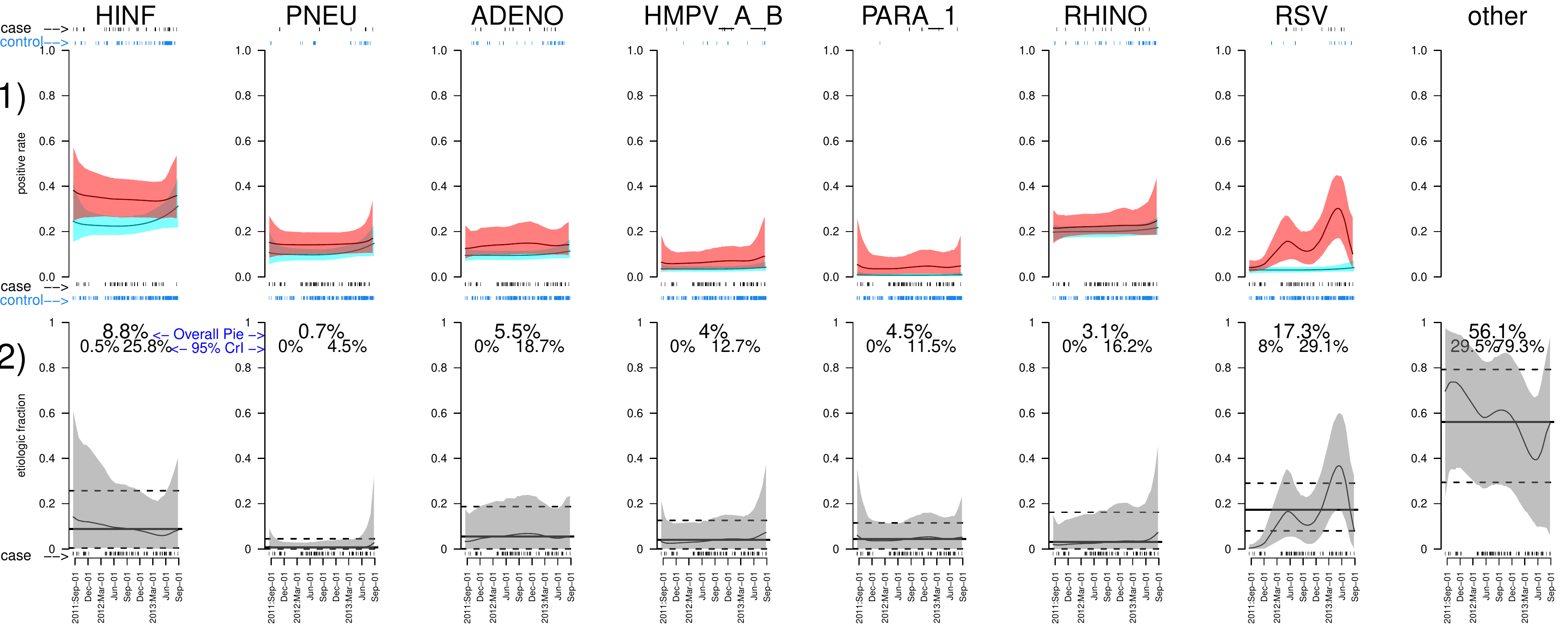}
\label{fig:older}
}
}
\addtocounter{figure}{0} 
\caption{Estimated seasonal PEF $\hat{\pi}_\ell({\sf date}, {\sf age}, {\sf severity}, {\sf HIV})$ for two most prevalent age-severity-HIV strata: \textbf{younger} (a) or \textbf{older} (b) than one, with severe pneumonia, HIV negative; Here the results are obtained from a model assuming seven single-pathogen causes ({\sf HINF}, {\sf PNEU}, {\sf ADENO}, {\sf HMPV.A.B}, {\sf PARA.1}, {\sf RHINO}, {\sf RSV}) and  an ``{\sf Not Specified}" cause. In an age-severity-HIV stratum and for each cause $\ell$:\\
Row 2) shows the temporal trend of $\hat{\pi}_\ell$ which is enveloped by pointwise $95\%$ credible bands shown in gray. The estimated overall PEF $\hat{\pi_\ell^\ast}$ averaged among cases in the present stratum is shown by a horizontal solid line, below and above which are two dashed black lines indicating the $2.5\%$ and $97.5\%$ posterior quantiles. The rug plot on the x-axis indicates cases' enrollment dates. \\
Row 1) shows the fitted temporal case (red) and control (blue) positive rate curves enclosed by the pointwise $95\%$ CrIs; The two rug plots at the top (bottom) indicate the dates of the cases and controls being enrolled and tested positive (negative) for the pathogen.}
\label{fig:data_results}
\end{figure}


The regression model accounts for stratification of etiology by the observed covariates and assigns cause-specific probabilities for two cases who have identical measurements but different covariate values.  Supplemental Figure S6 shows for two cases with all negative NPPCR results (the most frequent pattern among cases), the older case has a lower posterior probability of her disease caused by {\sf RSV} and higher probability of being caused by {\sf NoS}. Indeed, contrasting older and younger children while holding  the enrollment date, HIV, severity  constant, the estimated difference in the log odds (i.e., log odds ratio) of a child being caused by {\sf RSV} versus {\sf NoS} is negative: $-1.82~(95\%\text{ CrI}: -2.99,-0.77)$. 



Given age, severity and HIV status, we quantify the overall cause-specific disease burdens $\bpi^\ast$ by averaging the PEF function estimates by the empirical distribution of the enrollment dates. Contrasting the results in the two age-severity-HIV strata in Figure \ref{fig:younger} and \ref{fig:older}, since the case positive rate of {\sf RSV} among the older children reduces from $39.3\%$ to $17.9\%$ but the control positive rates remain similar (from $3.0\%$ to $4.1\%$), the overall PEF of {\sf RSV} ($\pi^\ast_{\tiny \sf RSV}$) decreases from $47.7 ~(95\%~\text{CrI}: 37.6,61.5)\%$ to $17.3~(95\%~\text{CrI}: 8.0, 29.1)\%$ and attributing a higher total fraction of cases to {\sf NoS} ($\pi^\ast_{\tiny \sf NoS}$) from $37.6~(95\%~\text{CrI}: 20.3,51.9)\%$ to $56.1~(95\%~\text{CrI}: 29.5,79.3)\%$; The overall PEFs for other causes remain similar.

\section{Discussion}
\label{sec::discussion}

In disease etiology studies where gold­-standard data are infeasible to obtain, epidemiologists need to integrate multiple sources of data of distinct quality to draw inference about the population and individual etiologic fractions. While the existing methods based on npLCM account for imperfect diagnostic sensitivities and specificities, complex measurement dependence and missingness, they do not describe the relationship between covariates and the PEFs. This paper addresses this analytic need by extending npLCM to a general regression modeling framework using case-control multivariate binary data to estimate disease etiology.  

The proposed approach has three distinguishing features: 1) It allows analysts to specify a model for the functional dependence of the PEFs upon important covariates. And with assumptions such as additivity, we can improve estimation stability for sparsely populated strata defined by many discrete covariates. 2) The model incorporate control data for the inference of PEF curve. The posterior inferential algorithm estimates a parsimonious covariate-dependent reference distribution of the diagnostic measurements from controls. Finally, 3) the model uses informative priors of the sensitivities (TPRs) only once in a population for which these priors were elicited. Relative to a fully-stratified npLCM analysis that reuses these priors, the proposed regression analysis avoids overly-optimistic etiology uncertainty estimates. 

We have shown by simulations that the regression approach accounts for population stratification by important covariates and as expected reduces estimation biases and produces $95\%$ credible intervals that have more valid empirical coverage rates than an npLCM analysis omitting covariates. In addition, the proposed regression analysis can readily integrate multiple sources of diagnostic measurements of distinct levels of diagnostic sensitivities and specificities, a subset of which are only available from cases (SS data), to further reduce the posterior uncertainty of the etiology estimates. Our regression analysis integrates the BrS and SS data from one PERCH site and reveals prominent dependence of the PEFs upon seasonality and a pneumonia child's age, HIV status and disease severity.

Future work may improve the proposed methods. First, flexible and parsimonious alternatives to the additive models may capture important interaction effects \citep[e.g.,][]{linero2018bayesian}. Second, in the presence of many covariates, class-specific predictor selection methods for $\pi_\ell(\bX_i)$ may provide further regularization and improve interpretability \citep{gustafson2008bayesian}. Third, when the subsets of pathogens that have caused the diseases in the population is unknown, the proposed method can be combined with subset selection procedures \citep{wu2019barlcm, gu2019learning}. Finally, scalable posterior inference for multinomial regression parameters \citep[e.g.,][]{zhang2017permuted} will likely improve the computational speed in the presence of a large number of disease classes and covariates.




\vspace{-.5cm}
\section*{Supplementary Materials}

\vspace{-.25cm}
The supplementary materials contain the technical details on prior specifications, a remark, additional simulation results and supplemental figures referenced in Main Paper.

\section*{Acknowledgment}

\vspace{-.25cm}
We thank the PERCH study team led by Kathernine O'Brien for providing the data and scientific advice, Scott Zeger, Maria Deloria-Knoll, Christine Prosperi and Qiyuan Shi for insightful comments and valuable feedback about \verb"baker" and Jing Chu for preliminary simulations. The research was partly supported by the Patient-Centered Outcomes Research Institute (PCORI) Award (ME-1408-20318, ZW), NIH grants P30CA046592 (National Cancer Institute Cancer Center Support Grant Development Funds, Rogel Cancer Center; ZW and IC), U01CA229437 (ZW) and an Investigator Award from Precision Health Initiative and an MCubed Award from University of Michigan (ZW).

\bibliography{nplcm-reg}
\bibliographystyle{apalike}

\newpage
\appendix

\begin{center}
{\large \bf Supplementary Materials}
\end{center}

The Supplementary Materials contain the technical details, a remark, extra simulation results and figures referenced in Main Paper. Section \ref{sec:prior_distribution} provides the technical specifications of the proposed prior distributions. Section \ref{sec:remark} remarks on model assumptions with covariates. Section \ref{sec:mcmc} details convergence checks for valid posterior inference. Section \ref{sec:simulation_main} presents more details of the simulations in Main Paper. Section \ref{sec:extra_simulation} presents additional simulation results. Finally, Section \ref{sec:supp_figures} contains Supplemental Figures.

\section{Prior distributions}
\label{sec:prior_distribution}

The unknown parameters include the regression coefficients in the etiology regression ($\{\bGamma_\ell^\pi\})$, the parameters in the subclass weight regression for the cases ($\{\bGamma_k^\eta\}$) and the controls  ($\{\bGamma^\nu_k\}$), the true and false positive rates $(\bTheta=\{\theta_{k}^{(j)}\}, \bPsi=\{\psi_{k}^{(j)}\})$. To mitigate potential overfitting and increase model interpretability, we \textit{a priori} place substantial probabilities on models with the following two features: (a) Few non-trivial subclasses via a novel additive half-Cauchy prior for the intercepts $\{\mu_{k0}\}$, and (b) for a continuous variable, smooth regression curves $\pi_\ell(\cdot)$, $\nu_k(\cdot)$ and $\eta_k(\cdot)$ by Bayesian Penalized-splines \citep[P-splines,][]{Lang2004} combined with shrinkage priors on the spline basis coefficients \citep{ni2015bayesian} to encourage towards constant values.

\subsection{Subclass Weight Regression: Encourage Few Subclasses}
\label{sec:prior.few.subclasses}

We propose a novel prior to encourage a small number of subclasses of non-trivial weights  in finite samples, or ``simplex regression shrinkage prior". We parameterize the intercepts $\{\mu_{k0}\}$ so that \textit{a priori} the higher-order subclasses are less likely to receive non-trivial weights. We let $\mu_{k0} = \sum_{j=1}^k u_{kj}\mu_{k0}^\ast$ where $u_{kj}, 1\leq j\leq k \leq K-1$ is a pre-specified triangular array of positive values. Upon heavy-tailed priors on $\mu_{k0}^\ast$ with positive supports, we will \textit{a priori} make higher-order subclasses increasingly less likely to receive substantial weights. In this paper, we use $u_{kj}  = 1, j=1, \ldots, k$; Other choices such as $u_{kj} = \ind\{k=j\}$ or $u_{kj} = 1/k$ may be useful in other settings. We specify the prior distributions of $\mu_{k0}^\ast$ to be heavy-tailed. In this paper we use Cauchy distribution with scale $s_0=10$. Since our control model take a classical latent class regression model form \citep{Bandeen1997} (the generic term ``class" here corresponds to control ``subclass" in an npLCM), the proposed prior for the subclass weight $\nu_k(\bW)=h_k(\bW; \bGamma^\nu_k), k=1, \ldots, K-1$ is also useful for a classical LCM regression analysis where the number of classes is unknown. Unlike a logistic stick-breaking specification $h_k(\bW; \cdot)$ without the intercepts $\{\mu_{k0}\}$, the proposed priors on the intercepts $\{\mu_{k0}\}$ encourage few subclasses and well recovers the true subclass weights. Using the same data simulated in Simulation I, Section 3 of Main Paper, Figure  \ref{fig:estimation_subwt} shows the proposed prior propagates into the posterior distribution and estimates 2 non-trivial subclasses from a working number of 7 subclasses.



At stick-breaking step $k$, the prior allows taking away nearly the entire stick segment currently left. Our basic idea is to have one of $\{g(\alpha_{ik})\}_{k=1}^{K-1}$ close to one \textit{a posteriori} by making the posterior mean of one of $\{\alpha_{ik}\}_{k=1}^K$ large. We accomplish this by designing a novel prior on the intercept $\mu_{k0} = \sum_{j=1}^k u_{kj}\mu_{k0}^\ast$ where
\begin{align*}
\mu^*_{k0}  \sim N^+(0, \tau_{0k}^{-1}), ~~
\tau_{0k} \sim {\sf Gamma}(a_{0}, b_{0}), k = 1, \ldots, K-1.
\end{align*}
The first level has a mean-zero Gaussian distribution truncated to the positive half. At the second-level, the precision (inverse variance) is Gamma distributed with shape $a_{0}=\nu/2$, and rate $b_{0}=\nu s_0^2/2$; it has the interpretation of $\nu$ prior independent sample(s) with a mean sample variance of $s_0^2$. Large values of $\tau^{-1}_{0k}$ help to stop stick-breaking at subclass $k$ forcing weights for ensuing subclasses $\nu_{k'}\approx 0$, $k'>k$, while small values let the stick-breaking scheme continue to step $k+1$. This type of prior sparsity, which we call ``selective stopping" or shrinkage over a simplex $\cS_{K-1}$ uniformly over covariates, effectively encourages using a small number of subclasses to approximate the observed $2^J$ probability contingency table for the control measurements in finite samples.


We accomplish selective stopping by the heavy right tail of $\mu^\ast_{k0}$'s marginal prior.  It has a truncated scaled-$t$ distribution with degree of freedom $\nu$ and scale $s_0$, and consequently peaks at zero and admits large positive values. Given other parameters in $\alpha^\nu_{ik}=\alpha^\nu_k(\bW_i; \bGamma_k^\nu)$, a near-zero intercept takes the stick-breaking procedure to the next step, while a large positive intercept effectively halts it. The tendency to stop at step $k$ is \textit{a priori} modulated by the scale parameter $s_0$.  Because, given the degree-of-freedom $\nu$, the prior probability $P(g(\alpha_{1k})>C\mid \nu, s_0), ~\forall C\in(0.5,1)$ approaches $1$  as the scale parameter $s_0$ increases.


In our simulations and applications,  we choose hyperparameters $\nu=1$ and $s_0=10$ for the intercept, and $k_{\beta}=4$ for the first B-spline coefficients $\bbeta_{kj}^{(1),\nu}$ in the prior (Equation \ref{eq:random.walk.prior}, Section \ref{sec:prior.smooth}). We have chosen our hyperparameters based on the interpretations on the probability (inverse-link) scale; see similar prior elicitations for regression coefficients in other applications \citep[e.g.,][]{bedrick1996new, witte1998software} and for automatic, stabilized and weakly-informative fitting of generalized linear models \citep{gelman2008weakly}. We choose the hyperparameters for the intercepts that put most prior mass of $g(\mu_{10})$ within $(0.5, 1-10^{-9})$, because $1-10^{-9}$ is sufficiently close to $1$ which means the stick-breaking is stopped at Step $k=1$. In contrast, we choose the first B-spline coefficient's hyperparameter $k_{\beta}=4$ that puts most prior mass of $g(\beta_{kj}^{(1),\nu})$ within $(0.02, 0.98)$, a range for the weight of a non-trivial subclass to break from the rest of the stick at Step $k$. Figure \ref{fig:hyperparameters} shows a sharp separation between the priors for $g(\mu^*_{k0})$ and $g(\beta_{kj}^{(1),\nu})$. The shapes of the priors again highlight the different roles played by the intercept and the B-spline coefficients: the former decides whether to continue the stick-breaking procedure to induce complex conditional dependence given covariates, and if so, the latter computes the fraction to break from the remaining length of the stick. The intercepts in the controls $\{\mu_{k0}\}$ are shared with the case subclass weight regression $\eta_k(\bW) = h_k(\bW; \bGamma^\eta_k)$; We set the same prior distributions for other elements of $\bGamma^\eta_k$, $k = 1, \ldots, K-1$.

\subsection{Encourage Smooth $f^\pi_{kj}$ and $f_{kj}$}
\label{sec:prior.smooth}

We use penalized B-splines to model the additive functions of a continuous variable in etiology regression  ($f^\pi_{kj}$), subclass weight regression for the cases and the controls ($f_{kj}$) \citep{Lang2004}. We expand $f^{\bullet}_{kj}(\cdot) = \sum_{c=1}^C\beta_{kj}^{(c)}B_{j}^{(c)}(\cdot)$, with $\{B_{j}^{(c)}(\cdot): c=1, \ldots, C\}$ being the shared $C$ cubic B-spline bases. We let $f^\pi_{\ell j}$, $f_{kj}$ in the case subclass weight regression and $f_{kj}$ in the control subclass weight regression have distinct coefficients: $\{\beta_{kj}^{(c),\pi}, k=1,\ldots, L\}$, $\{\beta_{kj}^{(c),\eta}, k=1,\ldots, K-1\}$ and $\{\beta_{kj}^{(c),\nu}, k=1,\ldots, K-1\}$, respectively.  With $M$ interior equally-spaced knots $\bkappa = (\kappa_0,\ldots, \kappa_{M+1})^\top$: $\min_i(x_{ij})=\kappa_0<\kappa_1<\cdots<\kappa_{M}<\kappa_{M+1}=\max_i(x_{ij})$, there are $C=M+4$ basis functions. It readily extends to let $f^\pi_{kj}$ and $f_{kj}$ have different numbers of basis functions.

Since the specification below applies to $f^\pi_{kj}(x; \bbeta_{kj}^\pi)$, $f_{kj}(w; \bbeta_{kj}^\nu)$ and $f_{kj}(w; \bbeta_{kj}^\eta)$ for any centered and standardized continuous variable, for simplicity we omit the superscripts $\pi, \nu, \eta$ and subscript $j$ .

The Penalized-splines in our formulation bypass the choice of the number and placement of knots $\bkappa$ by using a large number of knots deemed sufficient to capture the curves and imposing smoothing penalty on the coefficients for basis functions to prevent overfitting. The Gaussian random walk priors on basis coefficients are good choices for fitting Bayesian P-splines \citep{Lang2004}:
\begin{align}
\bbeta_{k} \mid \tau_{k}, \lambda_k \sim N(\bm{0}_{C\times 1}, \left(\tau_k\bm{K}\right)^{-1}),\label{eq:random.walk.prior}
\end{align}
where the symmetric penalty matrix $\bK=\Delta_1^\top\Delta_1$ is constructed from the first-order difference matrix $\Delta_1$ of dimension $(C-1)\times C$ that maps adjacent B-spline coefficients to $\beta_{k}^{(c)}-\beta_{k}^{(c-1)}$, $c=2, \ldots, C$ ({\small \verb"diff(diag(C),differences = 1)"} in R language), and $\tau_k$ is the smoothing parameters with large values leading to smoother fit of $f_{k}({x})$ (constant when $\tau_k=\infty$) and interpolation when near zero. This first-order random walk prior above uses a precision matrix $\bK$ of rank $C-1$ to model the adjacent differences. This leaves the prior of $\beta_{k1}$ unspecified, for which we further assign an independent prior $\beta_{k1}\sim N(0, k_{\beta}^{-1})$. We discuss the hyperparameter $k_{\beta}$ in the next subsection.

We use a mixture prior with two well-separated component distributions with one favoring small and the other large smoothing parameters $\tau_{kj}$:
\begin{align}
\tau_{kj} & \sim \xi_{kj}{\sf Gamma}(\cdot \mid a_\tau,b_\tau)+(1-\xi_{kj}){\sf InvPareto}(\cdot \mid a'_\tau, b'_\tau),\\
& {\sf InvPareto}(\tau ; a,b)  = \frac{a}{b}\left(\frac{\tau}{b}\right)^{a-1}, a>0, 0<\tau<b,
\end{align}
where the Gamma-distributed component ($a_\tau=3$, $b_\tau=2$) concentrates near smaller values while the inverse-Pareto component prefers larger values ($a'_\tau=1.5$, $b'_\tau=400$). This bimodal mixture distribution creates a sharp separation between flexible and smooth fits \citep{morrissey2011inferring, ni2015bayesian}. Because we use the first-order random walk prior, the most smooth fit is of degree $0$, i.e., constant functions. The random smoothness indicator $\xi_{kj}$ represents a flexible ($1$) or constant ($0$) shape of $f_k(\cdot)$. We let $\xi_{kj}\sim{\sf Bernoulli}(\rho)$ with success probability $\rho$ and then put a hyperprior $\rho\sim {\sf Beta}(a_\rho,b_\rho)$ to let data inform the degree of smoothness.



In this paper we use $a_\rho=0.5$, $b_\rho=1$ for each set of the B-spline basis coefficients for the cases ($\{\bbeta^{(c),\eta}_{kj}, c=1, \ldots, C\}$) and the controls ($\{\bbeta^{(c),\nu}_{kj}, c=1, \ldots, C\}$) to \textit{a priori}  give slight preference for constant curves, $k=1, \ldots, K-1$, $j=1, \ldots, q_1$; We use $a^\pi_\rho=1$, $b^\pi_\rho=0.5$ for the set of basis coefficients ($\{\bbeta^{(c),\pi}_{\ell j}, c=1, \ldots, C\}$) to \textit{a priori} give slight preference for flexible etiology regression functions, $\ell=1, \ldots, L$, $j=1, \ldots, p_1$. In the presence of high-dimensional covariates, the Beta prior with other hyperparameters can also allow a prior spread that lets the fraction of constant functions $\rho=\rho_p$ to approach $0$ as $p\rightarrow \infty$.

\subsection{Informative Prior Distributions for TPRs and FPRs}
\label{sec:tpr_priors}

The npLCM regression model is partially-identified \citep{Jones2010}. We assume independent informative priors for the TPRs in the BrS data likelihood: $\theta_k^{(j)}\overset{}{\sim}{\sf Beta}(a^{\sf \tiny BrS}_j,b^{\sf \tiny BrS}_j)$, $j=1, \ldots, J$, where ($a^{\sf \tiny BrS}_j$,$b^{\sf \tiny BrS}_j$) are chosen so that the $2.5\%$ and $97.5\%$ quantiles match a prior range elicited from laboratory scientists \citep{deloria2017bayesian}. In the presence of SS data for a subset of pathogens (e.g., culturing bacteria from blood), we similarly set the hyperparameters for the Beta distribution of the TPRs of the SS measures  where ranges can be computed from existing vaccine probe trials \citep[e.g.,][]{feikin2014use}. Since the control data provide direct estimates of the FPRs, we specify independent priors for $\psi_k^{(j)}\sim {\sf Beta}(1,1), j=1,\ldots, J, k=1,\ldots, K$.


\section{Remark on the Control Model with Covariates}
\label{sec:remark}

The proposed model for the control data with covariates $\bW$ is a generative model where we first draw a subclass indicator $Z \mid \bW \sim  {\sf Categorical}_K\{\bnu(\bW)\}$, and generate measurements $M_{j}\mid Z=k$ according to a Bernoulli distribution with positive rate $\psi_k^{(j)}$, independently for $j=1,...,J$. By assuming mutually independent measurements $M_{1}, \ldots, M_{J}$ given subclass $Z$ and $Y=0$, we let the covariates influence the dependence structure of the measurement only through the unobserved $Z$. As a result, upon integrating over $Z$, the proposed model does \textit{not} assume marginal independence $\PP(\bM\mid \bW, Y=0) = \prod_{j=1}^J\PP(M_j\mid \bW,Y=0)$ in contrast to a kernel-based extension of the pLCM that makes this assumption \citep[][Supplementary appendix]{saha2018causes}. Our approach to incorporating covariates to model control data follows \citet{Bandeen1997}; For other approaches, see examples in the study of particulate matter \citep{gryparis2007semiparametric}, HIV population size estimation \citep{bartolucci2006class}, and alcoholic and drug addiction \citep{chung2006latent}. 

\section{Convergence Checks}
\label{sec:mcmc}
In simulations and data analysis, we ran three MCMC chains each with a burn-in period of $10,000$ iterations followed by $10,000$ iterations stored for posterior inference. We look for potential non-convergence in terms of Gelman-Rubin statistic \citep{brooks1998general} that compares between-chain and within-chain variances for each model parameter where a large difference ($R_c>1.1$) indicates non-convergence; We also used Geweke's diagnostic \citep{geweke1996measuring} that compare the observed mean for each unknown variable using the first $10\%$ and the last $50\%$ of the stored samples where a large $Z$-score indicates non-convergence ($|Z|>2$). In our simulations and data analyses, we observed fast convergence (many satisfied convergence criteria within $2,000$ iterations) that led to well recovered regression curves, TPRs and FPRs.

\section{Additional Information about Simulations of Main Paper}
\label{sec:simulation_main}

\noindent \underline{\it Simulation {\sf I}.} we let $\pi_\ell(\cdot)$, $\nu_k(\cdot)$ and $\eta_k(\cdot)$ depend on the two covariates $\bX=\bW=(S,T)$, $S$ and enrollment date ($T$), so that regression adjustments are necessary (see Remark 1, Main Paper). We simulate BrS measurements on $J=9$ pathogens and assume the number of potential single-pathogen causes $L=J=9$. To specify etiology regression functions that satisfy the constraint $\sum_{\ell=1}^L\pi_\ell(\bx) = 1$, we use stick-breaking parameterization with $L=9$ segments. In particular, we let $\logit~\{ g_1(s,t)\} = \beta_{1}\ind(s=1)+\sin(8\pi(t-0.5)/7)$, $\logit~\{ g_2(s,t)\} = \beta_{2}\ind(s=1)+4\exp(3t)/(1+\exp(3t))-0.5$, $\logit(g_\ell)= \beta_{8}\ind(s=1)$ for $\ell>2$; Let the PEF functions $\pi_\ell(s,t) = g_\ell(s,t)\prod_{j< \ell}\{1-g_j(s,t)\}, \ell = 1, \ldots, L(=9)$, where $\beta_{\ell}=0.1, \ell=1, \ldots, 8$. The true control distribution depend on covariates with $K=2$ subclass weight functions:
$\nu_1(s,t) = \logit^{-1} \left\{\gamma^\nu_{1}\ind(s=1)+4\exp(3t)/(1+\exp(3t))-0.5\right\}$ and $\nu_2(s,t) = 1-\nu_1(s,t)$. We specify $\eta_k(s,t)=\nu_k(s,-t), k=1,2$, highlighting the need for using different subclass weights among cases and controls in an npLCM analysis. We set the true TPRs $\theta_k^{(j)} = 0.95$ and the FPRs $\psi_1^{(j)}=0.5$ and $\psi_2^{(j)}=0.05$.

In the regression analyses, we set $\phi_\ell(\bX)$ to be an additive model of a $\ind\{S=2\}$ indicator and a B-spline expansion with  $7$ degrees of freedom (d.f.) for standardized enrollment date $t$. We use $K^*=7$ and specify the regression formula for subclass weights $\nu_k(\cdot)$ and $\eta_k(\cdot)$ by additive models of the $\ind\{S=2\}$ indicator and a B-spline expansion with 5 d.f. for standardized enrollment date. 

 \noindent \underline{\it Simulation {\sf II}.} We consider $L=J=3,6,9$ causes, under single-pathogen-cause assumption, BrS measurements made on $N_d$ cases and $N_u$ controls for each level of $X$ where $N_d=N_u=250$ or $500$. The functions $\phi_\ell(X) = \beta_{0\ell}+\beta_{1\ell}\ind\{ X=2\}$ take two sets of values to reflect how variable the PEFs are across the two $X$ levels: i) $\bbeta^{\sf i}_0=(0,0,0,0,0,0)$ and $\bbeta^{\sf i}_1=(-1.5,0,-1.5,-1.5,0,-1.5)$ where causes have uniform PEFs when $X=1$ and causes {\sf B} and {\sf E} dominate when $X=2$, or ii) $\bbeta^{\sf ii}_0=(1,0,1,1,0,1)$ and $\bbeta^{\sf ii}_1=(-1.5,1,-1.5,-1.5,1,-1.5)$ to mimic the scenario where pathogens {\sf B} and {\sf E} have lower PEFs when $X=1$ and occupy more fractions when $X=2$. We further let the measurement error parameters take distinct values of the TPRs $\theta_k^{(j)}=0.95$ or $0.8$ and the FPRs $(\psi_1^{(j)}, \psi_2^{(j)}) \in \{(0.5,0.05), (0.5,0.15)\}$,  for $j=1, \ldots, J$. Finally, we set the truth $\nu_k(W) = \eta_k(W)  =\logit^{-1} \left(\gamma_{k0}+\gamma_{k1} \ind\{W=2\}\right)$ where
$(\gamma_{10},\gamma_{11})=(-0.5,1.5)$ and $(\gamma_{20},\gamma_{21})=(1,-1.5)$. 

\noindent \underline{\it Simulation {\sf II}: a randomly chosen replication.} Here we illustrate the inferences about the stratum-specific and overall PEFs that are available to an analyst by considering a two-level covariate $X=W$ with $J=6$ measurements. Under the single-pathogen cause assumption, we can estimate $12=(2\times 6)$ PEFs, six per level of $X$ as well as six overall PEFs. For example, based on a single data set simulated under the scenario $\{L=6$, $N_d = 500$, $K=2$, $\theta_k^{(j)}=0.8$, $(\psi_1^{(j)}, \psi_2^{(j)})=(0.5,0.05)$, ($\bbeta^{\sf ii}_0, \bbeta^{\sf ii}_1)\}$, Supplemental Figure \ref{fig:covariate_adjusted_etiology} shows the posterior distribution of the stratum-specific etiology fractions $\pi_{\ell}(X=s)$ for ($s=1,2$) by row and $L(=J)$ causes $(\ell=1, \ldots, 6)$ by column with the true values indicated by the blue vertical dashed lines; The bottom row shows the posterior distribution of $\pi_\ell^\ast=\sum_s w_s \pi_\ell(X=s)$ for $L$ causes with empirical weights $w_s=N_d^{-1}\sum_{i:Y_i=1}\ind\{ X_i=s\}$, $s=1,2$. The true stratum-specific and overall PEFs are covered by their respective $95\%$ CrIs.

\section{Additional Simulation Results}
\label{sec:extra_simulation}
\subsection{Estimating $\pi_\ell(X)$}
\label{sec:estimate_pix}
We use simulation studies to show the frequentist performance of the npLCM regression model in recovering stratum-specific PEFs; The results below are based on a single discrete covariate that influence the PEFs but not the subclass weights in the cases or controls.

In this simulation study, we simulate $500$ cases and $500$ controls for each of $7$ sites. Every subject is measured on $6$ pathogens {\sf A} to {\sf F}; The causes of disease are single-pathogen causes {\sf A}-{\sf F}. First, we let the PEFs vary by site which are shown in Table \ref{etiology-truth}. Second, we simulate the data using $K=1$ subclass.

\begin{table}[H]
\begin{center}
\captionsetup{width=0.85\linewidth}
\caption{True PEFs for seven sites (boldfaced numbers indicate the highest PEFs within each stratum).}
 \begin{tabular}{c c c c c c c} 
 \hline
site\textbackslash cause  & A &  B &   C &   D &  E &  F \\ [0.5ex] 
 \hline
1 & \textbf{0.5}&0.2&0.15&0.05&0.05&0.05\\ 
2 & 0.2&\textbf{0.5}&0.15&0.05&0.05&0.05\\
3 & 0.2&0.15&\textbf{0.5}&0.05&0.05&0.05 \\
 4 & 0.2&0.15&0.05&\textbf{0.5}&0.05&0.05\\
  5 & 0.2&0.15&0.05&0.05&\textbf{0.5}&0.05\\
6 & 0.2&0.15&0.05&0.05& 0.05&\textbf{ 0.5}\\ 
7 &0.05&0.2&0.15&\textbf{0.5}&0.05&0.05\\
 \hline 
\end{tabular}
\label{etiology-truth}
\end{center}
\end{table}

We simulate data under two TPR scenarios {\sf (I)} strong signal with $\theta_1^{(j)}=0.99$ and $\psi_1^{(1)}=0.01$ where data are expected to provide strong information about the PEFs, and {\sf (II)} weak signal with $\theta_1^{(j)}=0.55$ and $\psi_1^{(1)}=0.45$ where it is easy to confuse true and false positive results and the data do not provide strong information about the PEFs. In both scenario {\sf (I)}  and {\sf (II)} , we used a Beta(6,2) distribution as a prior for the TPRs of the BrS measurements. We set the true TPRs and FPRs to be the same across sites and pathogens. In fitting the regression models, we use the etiology regression formulation by specifying $L-1$ sets of regression parameters with site dummy variables as the predictors in $\phi_\ell(\cdot)$. Since our goal is to infer $S=7$ sets of PEFs, we can also specify $S=7$ sets of symmetric Dirichlet priors with hyperparameter $\alpha$ ({\sf Dir}($\alpha$)); We use $\alpha=1$ here. The package \verb"baker" (\url{https://github.com/zhenkewu/baker}) provides an option to use Dirichlet priors when the PEFs depend on discrete covariates only.

\subsubsection{Scenario I: Strong Signal}
\label{question-2}

Over $R=100$ replications, the top half  of Table \ref{table:simulation_appendix} summarizes the coverage rates of the $95\%$ credible intervals (CrIs) for the PEFs across all the sites. We observed excellent recovery of the true values across all causes and sites with the $95\%$ CrIs covered the true values between $90\%$ to $100\%$ of the time. Panel I of Table \ref{table:simulation_appendix} also shows for site 1 the posterior mean PEFs, posterior standard deviations (sd's) of the PEFs, and posterior mean squared errors (PMSEs, estimated by $B^{-1}\sum_{b=1}^B\sum_{i: Y_i=1}\{\pi_\ell(X_i=s; \bgamma^{\pi,(b)})-\pi_\ell^{0}(X_i=s)\}$ with $B$ retained posterior samples $\{\bgamma^{\pi,(b)}\}$) averaged over $R$ replications. The posterior means provide excellent estimation of the PEFs with small average PMSEs.

\subsubsection{Scenario II: Weak Signal}
\label{similar}

Using data simulated under less discrepant TPRs and FPRs than those in Scenario {\sf I}, the $95\%$ CrIs cover the truths well for most site-cause pairs, but undercover the truths for causes with the highest PEF in each site (see Table \ref{table:II-wideTPR-uniformDir-cover}). This is expected because when the signal from the data is weak, the model relies more heavily on the uniform prior distribution for the PEFs (symmetric Dirichlet prior with hyper-parameter $1$). 

\begin{table}[!h]
\begin{center}
\captionsetup{width=0.6\linewidth}
\caption{Number of times (out of 100 replications) that the true value is covered by the $95\%$ CrIs (Scenario II, {\sf Beta}(6,2) prior for the TPRs). Boldfaced numbers indicate the highest PEFs ($0.5$) within each stratum.}
 \begin{tabular}{c c c c c c c} 
 \hline
site\textbackslash cause & A &  B &  C &  D & E &  F \\ [0.5ex] 
 \hline
1 &  \textbf{73} & 100 & 100  & 99 & 100 & 100 \\ 
2 & 100 & \textbf{79} & 100 & 100 & 100  & 99\\
3 &  100 & 100 & \textbf{83}&  98 & 100 & 100 \\
 4 &  100 & 100 & 100 &  \textbf{73} & 100 &  99 \\
 5 &  99 & 100 & 100 & 100 &  \textbf{85} & 100 \\
6 & 100 & 100  &100  & 99 & 100  & \textbf{88}\\ 
7 & 100 & 100 & 100 &  \textbf{81} & 100 &  99 \\ 
 \hline 
\end{tabular}
\label{table:II-wideTPR-uniformDir-cover}
\end{center}
\end{table}

\noindent \textit{More Informative TPR Priors (II$^\ast$).} We further investigate the model performance when we change the TPR prior distributions from the {\sf Beta}(6,2) to a Beta distribution that has 95\% of its mass between 0.525 and 0.575 and is around the true TPRs ({\sf Beta}(835.95, 683.79); \verb"beta_parms_from_quantiles(c(0.525,0.575))" using \verb"baker"). Panel $II^\ast$ of Table \ref{table:simulation_appendix} shows dramatic improvements in the coverage rates. These results suggest that changing the prior distributions of the TPRs so that it is more tightly concentrated around plausible values can improve inferences of the stratum-specific PEFs in the presence of high levels of noises. Relative to Scenario {\sf I}, the average PMSEs are larger across sites and pathogens reflecting the weaker signal in this setting.

In summary, in the simulation study where the PEFs are influenced by a discrete covariate, the regression model recovers the true values well under high signals (high sensitivities and low FPRs). Under lower sensitivities and higher FPRs, the noisier simulated data are less informative about the PEFs which are then more influenced by the prior distributions of the TPRs and PEFs. In practice, we recommend eliciting quality informative TPR priors from domain scientists as in the PERCH study and perform sensitivity analyses to understand the robustness of the results with respect to the prior distributions.

\begin{table}[h]
\caption{Scenario I and II$^\ast$: coverage rates of the $95\%$ CrIs; For Site 1, the posterior means, standard deviations (s.d.'s) and PMSE of the stratum-specific PEFs averaged over $R=100$ replications are also shown. Boldfaced numbers indicate the highest PEFs ($0.5$) within each stratum.}
\begin{tabular}{llccccccc}
\cline{1-9}
&  & site  \textbackslash cause & {\sf A}   & {\sf B}   & {\sf C} & {\sf D} &{\sf E} & {\sf F} \\ \cline{2-9} 
\multirow{12}{*}{I} & \multirow{7}{*}{coverage}  & 1 &  \textbf{99}  & 93  & 97  & 94  & 96 &  90 \\ 
& & 2 &  97 &   \textbf{90} &   96  &  97 &   95 &   94\\
& & 3 & 100  &  95  & \textbf{ 98}  &  98  &  95  &  96 \\
& &  4 & 93 &   94 &   96  &   \textbf{95 } &  92 &   99 \\
& &   5 & 96 &   94  &  96 &   97 &   \textbf{95} &   98\\
& & 6 & 96  &  97  &  98 &   99 &   95 &   \textbf{96} \\ 
& & 7 & 96  &  97 &   91 &   \textbf{100} &   95 &   96 \\ \cline{2-9}
& \multirow{4}{*}{\begin{tabular}[c]{@{}l@{}}posterior\\ summary\end{tabular}} & truth (\underline{Site 1})  & {0.5} & {0.2} & {0.15} & {0.05} & {0.05} & {0.05} \\
& &    average of post. mean &  0.495 & 0.197 & 0.152 & 0.053 & 0.053 & 0.051 \\
& &    average of post. s.d. &  0.023 & 0.018 & 0.016 & 0.01 & 0.01 & 0.01 \\
& &    average PMSE &   0.0010 & 0.0007 &  0.0005 &  0.0002  & 0.0002 &  0.0002 \\ \cline{1-9} 
\multirow{12}{*}{II$^\ast$} & \multirow{7}{*}{coverage}  & 1   &  \textbf{98}  & 89  & 98   & 99   & 100  & 100  \\
&  & 2   & 97  &  \textbf{95}  & 96   & 100  & 100  & 99   \\
&  & 3   & 93  & 98  &  \textbf{91}   & 99   & 99   & 100  \\
&  & 4   & 95  & 98  & 100  &  \textbf{95}   & 99   & 100  \\
&  & 5   & 94  & 94  & 99   & 99   &  \textbf{91}   & 100  \\
&  & 6   & 95  & 97  & 100  & 99   & 99   &  \textbf{90}   \\
&  & 7   & 100    & 95  & 94   & \textbf{96}   & 100  & 99   \\ \cline{2-9} 
& \multirow{4}{*}{\begin{tabular}[c]{@{}l@{}}posterior\\ summary\end{tabular}} & truth (\underline{Site 1})  &  {0.5} &  {0.2} &  {0.15} &  {0.05} &  {0.05} &  {0.05} \\
&  & average post. mean & 0.417  & 0.163  & 0.138   & 0.091   & 0.086   & 0.106   \\
&  & average post. s.d. & 0.27   & 0.174  & 0.162   & 0.135   & 0.13    & 0.141   \\
&  & average PMSE   & 0.131  & 0.067  & 0.056   & 0.034   & 0.031   & 0.042   \\ \cline{1-9} 
\end{tabular}
\label{table:simulation_appendix}
\end{table}

\subsection{Valid inference of $\pi^\ast_\ell$ omitting covariates}
\label{sec:valid_though_nox}

Under assumption (A1) in Remark 1 of Main Paper, the case subclass weights $\bEta_k(\bW)=\eta_k$, $k=1, \ldots, K$, we conduct a simulation study to show that an npLCM analysis omitting covariates is able to provide valid inference about the overall PEFs ($\bpi_\ell^\ast$). The simulation settings are exactly the same as in Simulation II, Section 4 of Main Paper, except that we set $\gamma_{20}=\gamma_{21}=0$ to satisfy assumption (A1). Figure \ref{fig:noXFPR_res_relbias} shows the percent relative biases are similarly negligible in all the $16$ scenarios with $6$ disease classes; Figure \ref{fig:noXFPR_res_coverage} shows excellent empirical coverage rates of the $95\%$ CrIs for $\{\pi_\ell^\ast\}$.

\newpage
\section{Supplemental Figures}
\label{sec:supp_figures}

%



\begin{figure}[H]
\centering
\includegraphics[width=.6\textwidth]{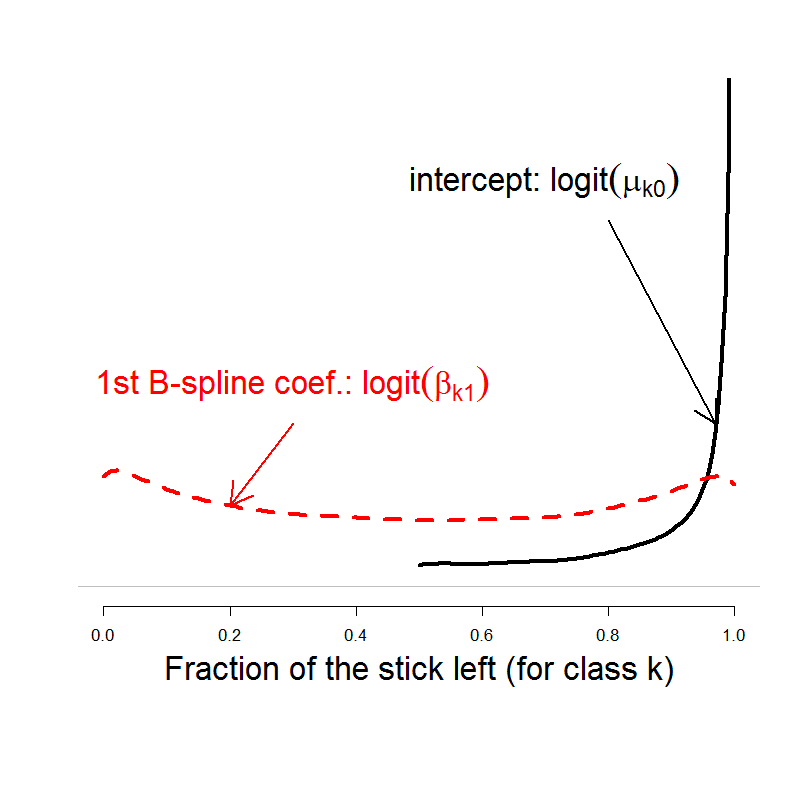}
\caption{\footnotesize Prior densities for $\logit(\alpha^\nu_{ik})$, the fraction to be broken for subclass $k$ from the stick currently left,  when $\alpha_{ik}$ equals: 1) the intercept $\mu^\ast_{k0}$ (\textit{black, solid line}) or 2) the first B-spline coefficient $\beta_{kj}^{(1),\nu}$ (\textit{red, broken line}). The former concentrates near $1$ because $\mu^*_{k0}$ has a scaled-$t$ distributed prior that puts substantial mass at the right tail; much less so for the latter.}
\label{fig:hyperparameters}
\end{figure}

\begin{figure}[H]
\captionsetup{width=\linewidth}
\centering
\addtocounter{figure}{0} 
\raisebox{-.5in}{
\subfigure[case]{
 \includegraphics[width=0.9\linewidth]
{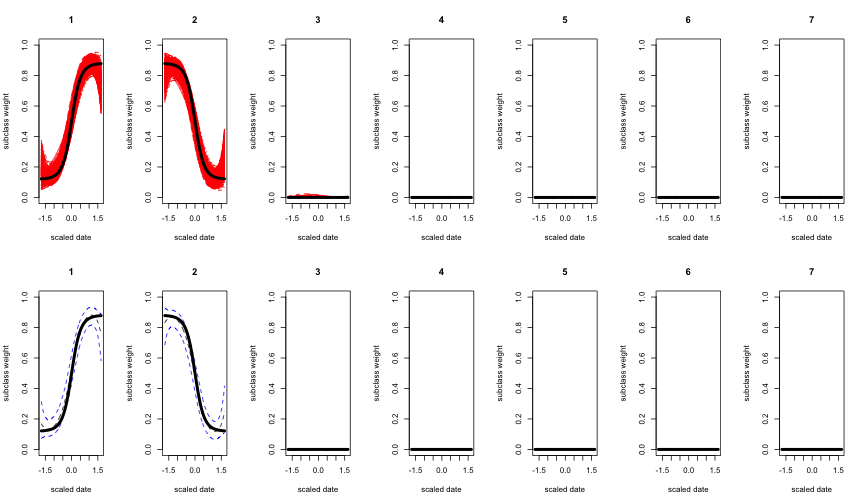}
\label{fig:case_subwt}
}
}
\hspace*{.2in}
{\subfigure[control]{
\includegraphics[width=0.9\linewidth]{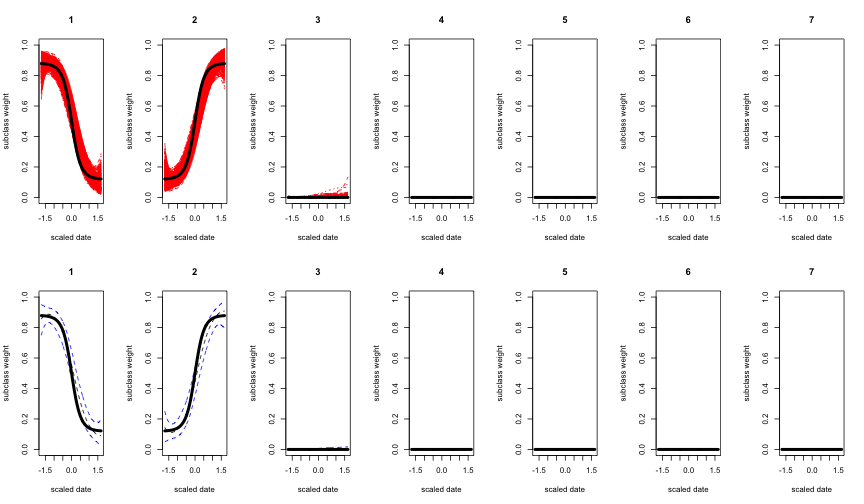}
\label{fig:ctrl_subwt}
}
}
\addtocounter{figure}{0} 
\caption{\footnotesize By propagating the prior that encourages few subclasses, the algorithm correctly infers two subclasses from the simulated data in Simulation I, Section 4 of Main Paper. Estimated case (top) and control (bottom) subclass weight curves for seven subclasses over one continuous covariate $\widehat{\nu}_k(t)$ (central blue dashed lines enclosed by the $95\%$ credible regions; the red curves are posterior samples) compared against the simulation truths ($\nu^0_k(t)$, black solid lines). The number of subclasses is bounded by seven during model fitting.}
\label{fig:estimation_subwt}
\end{figure}

\begin{figure}[H]
\centering
\includegraphics[width=\textwidth]{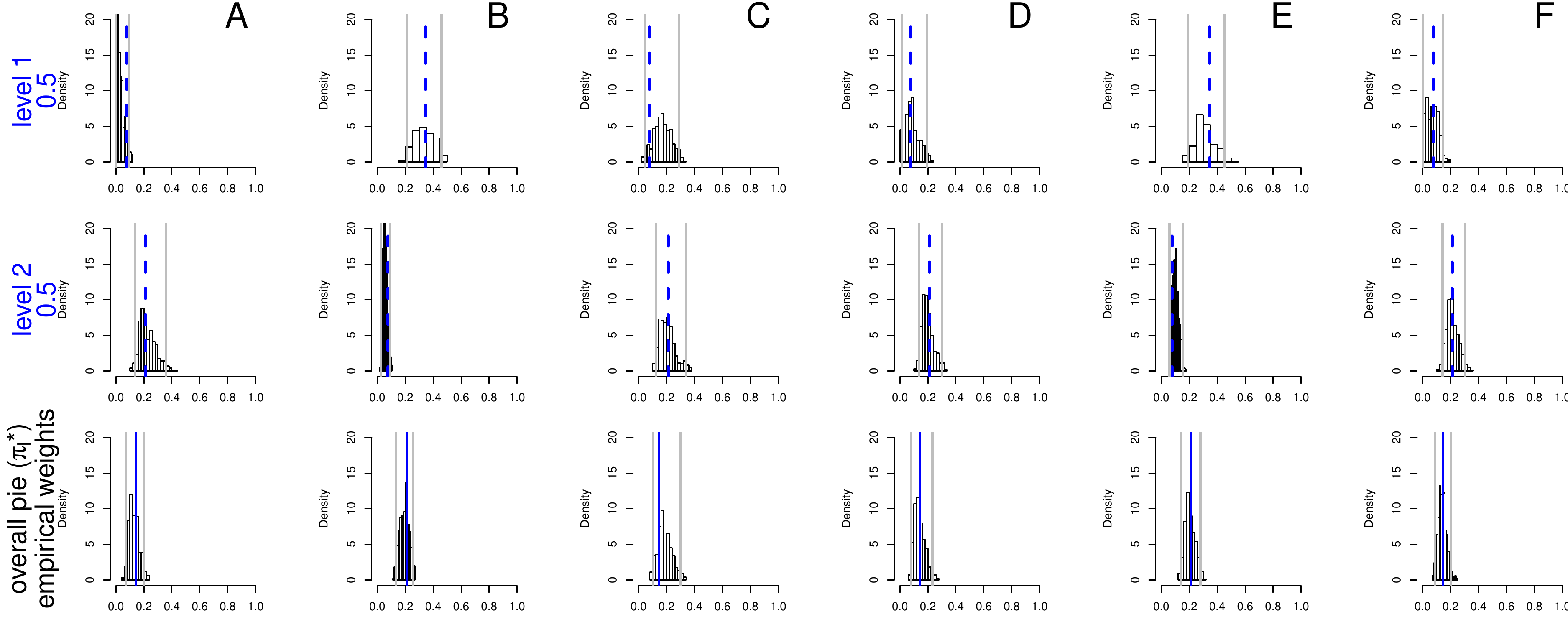}
\caption{\footnotesize Posterior distributions of the stratum-specific (Row 1 and 2) and the overall (Bottom Row) PEFs based on a simulation with a two-level discrete covariate and $L=J=6$ causes. The vertical gray lines indicate the $2.5\%$ and $97.5\%$ posterior quantiles, respectively; The truths are indicated by vertical blue dashed lines. {\it Row 1-2}) PEFs by stratum ({\sf level = 1,2}) and cause ({\sf A-F}); \textit{Bottom}) $\pi^\ast_\ell$: overall population etiologic fraction for cause {\sf A-F} (empirical average of the two PEFs above).}
\label{fig:covariate_adjusted_etiology}
\end{figure}

\begin{figure}[H]
\captionsetup{width=.9\linewidth}
\centering
\addtocounter{figure}{1} 
\raisebox{-.5in}{
\subfigure[]{
 \includegraphics[width=0.9\linewidth]
{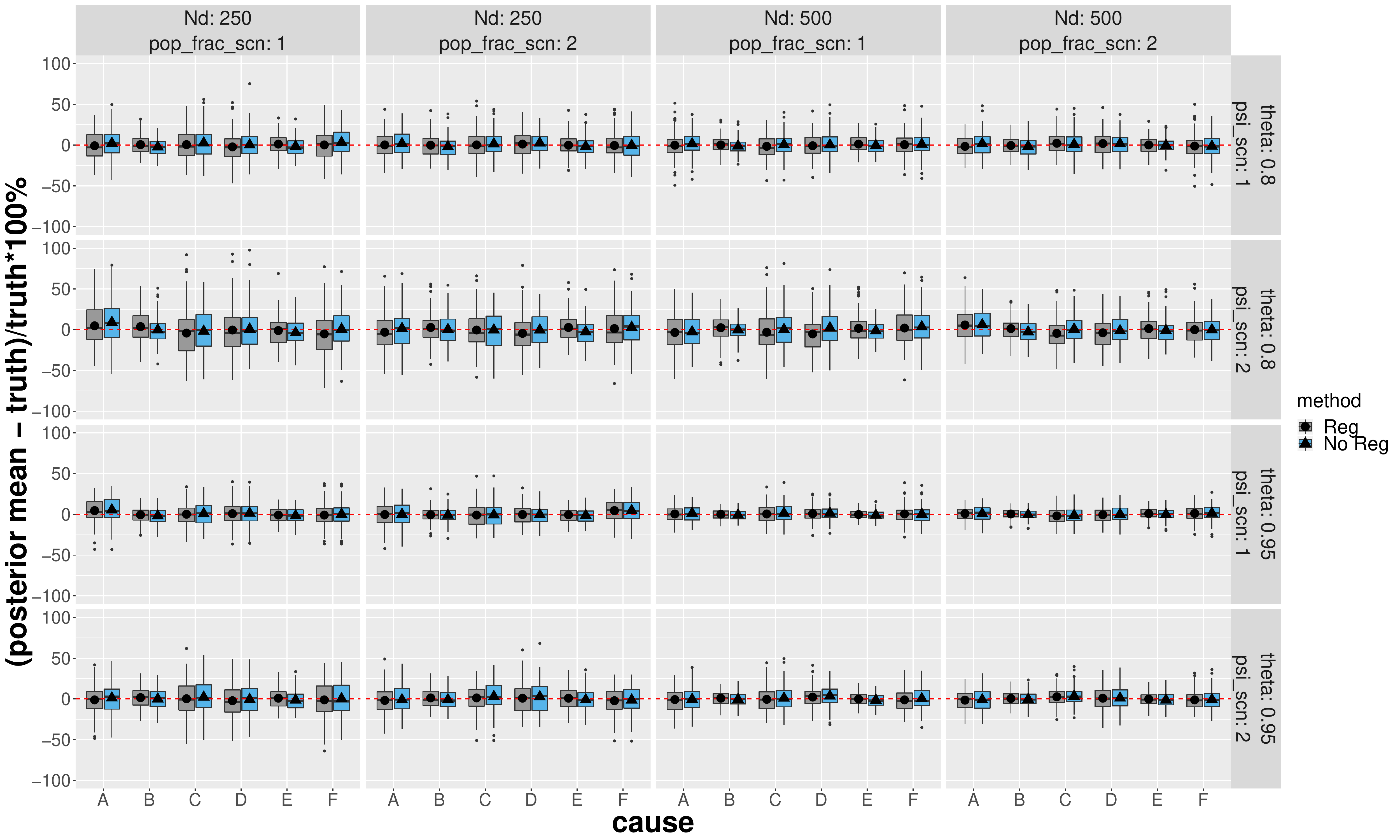}
\label{fig:noXFPR_res_relbias}
}
}
\hspace*{.2in}
{\subfigure[]{
\includegraphics[width=0.9\linewidth]{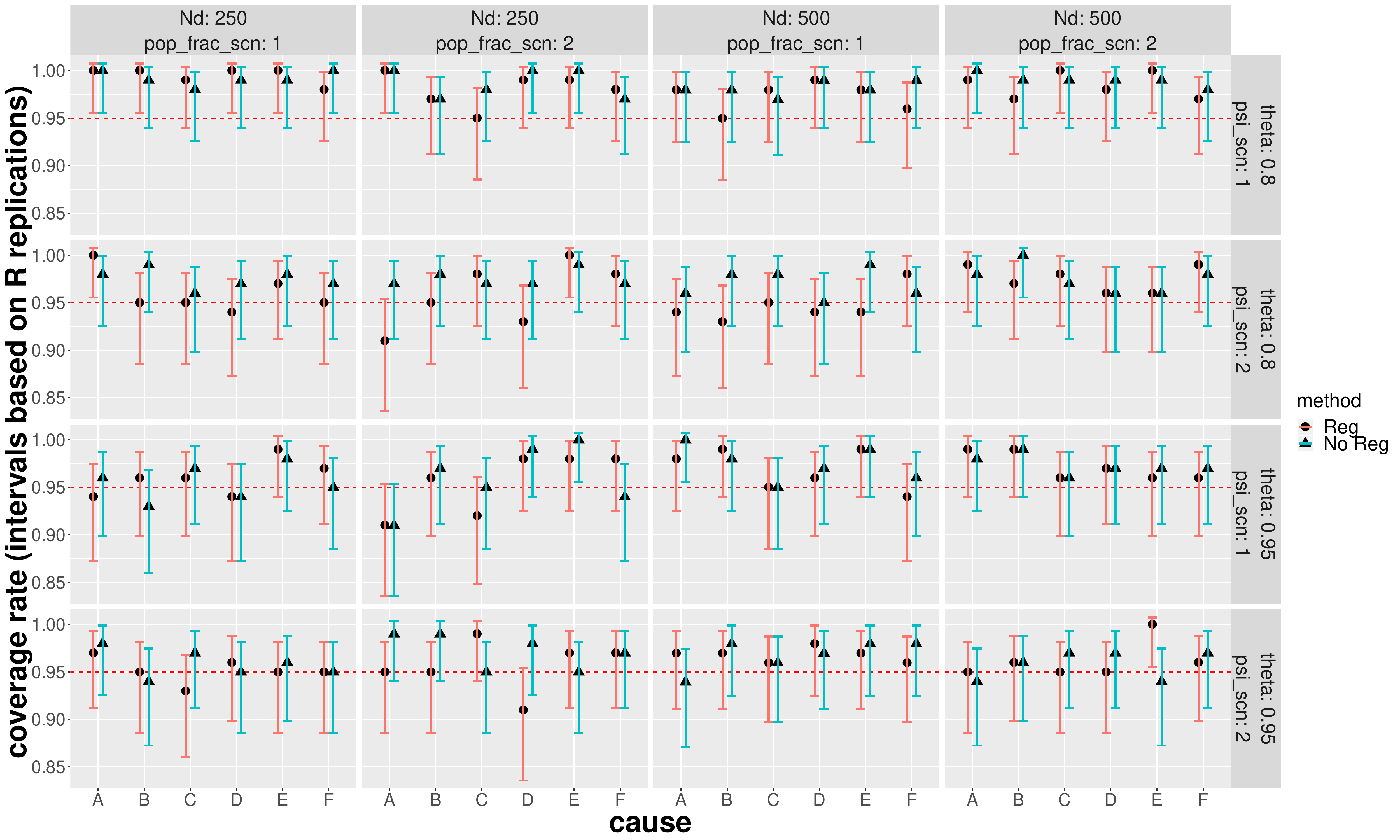}
\label{fig:noXFPR_res_coverage}
}
}
\addtocounter{figure}{-1} 
\caption{\footnotesize NPLCM analyses with or without regression perform similarly in terms of percent relative bias (top) and empirical coverage rates (bottom) over $R=100$ replications in simulations where the case and control subclass weights \textit{do not} vary by covariates.  Each panel corresponds to one of $16$ combinations of true parameter values and sample sizes. See Figure 3 in Main Paper for detailed descriptions of the figure.}
\label{fig:similar_overall_pie}
\end{figure}

\begin{landscape}
\begin{figure}[H]
\centering
\includegraphics[width=.88\linewidth]{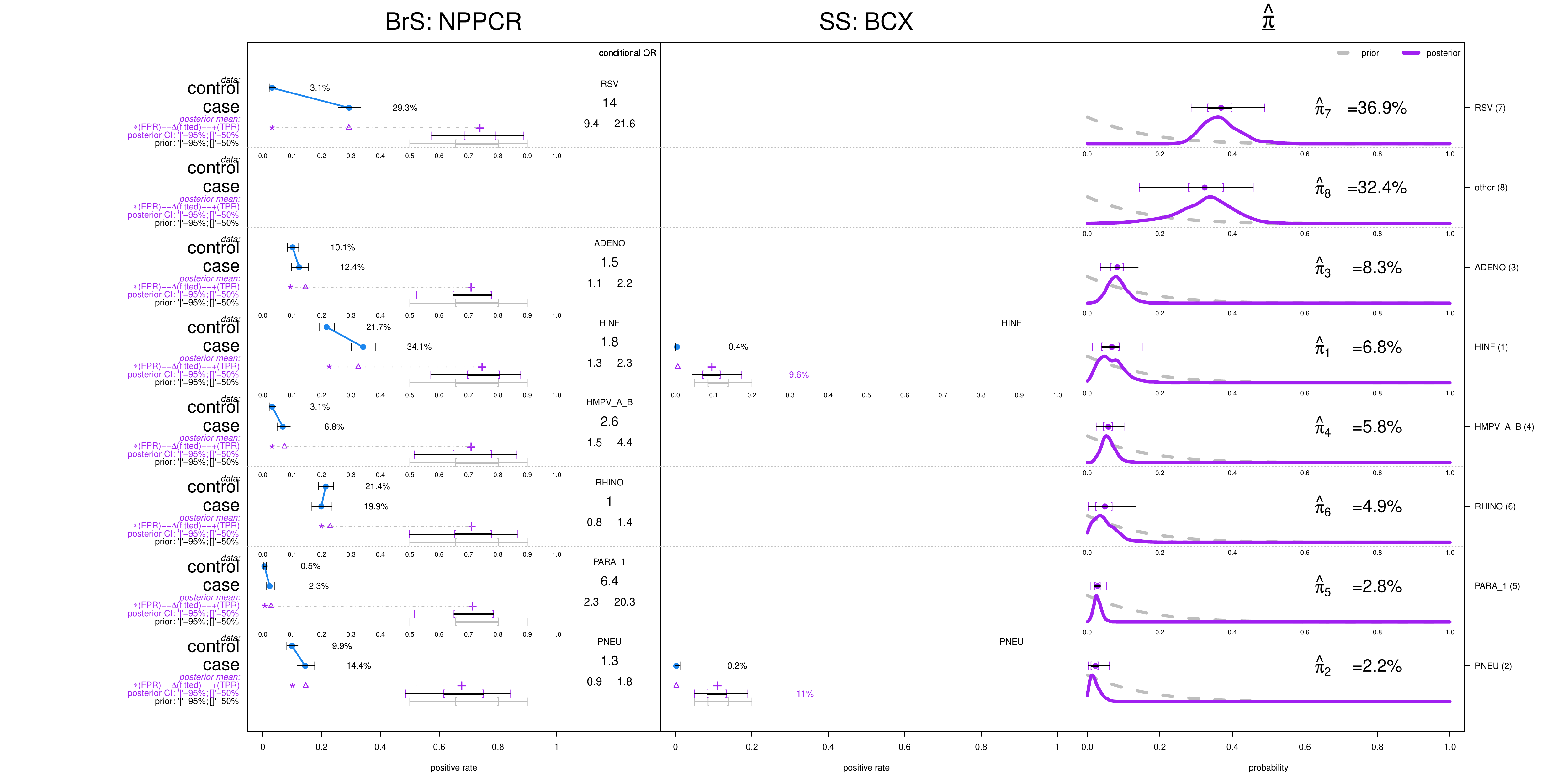}
\caption{\footnotesize Panel plot with BrS, SS and Etiology Pies obtained from an npLCM analysis omitting covariates ($K=5$). For each of the 7 pathogens, a summary of the BrS and SS data analyzed in Section 5 of Main Paper is shown in the left two columns, along with some of the intermediate model results; and the prior and posterior distributions for the PEFs on the right (rows ordered by posterior means).  \textit{Left}) The observed BrS rates (with $95\%$ confidence intervals, CI) for cases and controls are shown on the far left with solid dots. The conditional odds ratio (COR) contrasting the case and control rates given the other pathogens is listed with $95\%$ CI in the box to the right of the BrS data summary. Below the case and control observed rates is a horizontal line with a triangle. From left to right, the line starts at the estimated false positive rate (FPR, $\widehat{\psi}_j^{\BrS}$) and ends at the estimated true positive rate (TPR, $\widehat{\theta}_j^{\BrS}$), both obtained from the model. Below the TPR are $95\%$ and $50\%$ intervals summarizing its posterior (top) and prior (bottom) distributions for that pathogen. These intervals show how the prior assumption influences the TPR estimate as expected given the identifiability constraints. The triangle on the line is the model estimate of the case rate to compare to the observed value above it. \textit{Middle}) The SS data are shown in a similar fashion to the right of the BrS data. By definition, the FPR is 0.0 for SS measures and there is no control data. The observed rate for the cases is shown with its $95\%$ CI. The estimated SS TPR ($\widehat{\theta}^{\SSs}_j$) with prior and posterior distributions is shown as for the BrS data. \textit{Right}) The marginal posterior and prior distributions of the etiologic fraction for each pathogen. We appropriately normalized each density to match the height of the prior and posterior curves. The posterior mean with $50\%$ and $95\%$ CrIs are shown above the density.
}
\label{fig:hyperparameters}
\end{figure}
\end{landscape}

\begin{figure}[H]
\captionsetup{width=0.9\linewidth}
\centering
\addtocounter{figure}{0} 
\raisebox{-.0in}{
\subfigure[Cause: {\sf RSV}]{
 \includegraphics[width=0.45\linewidth]
{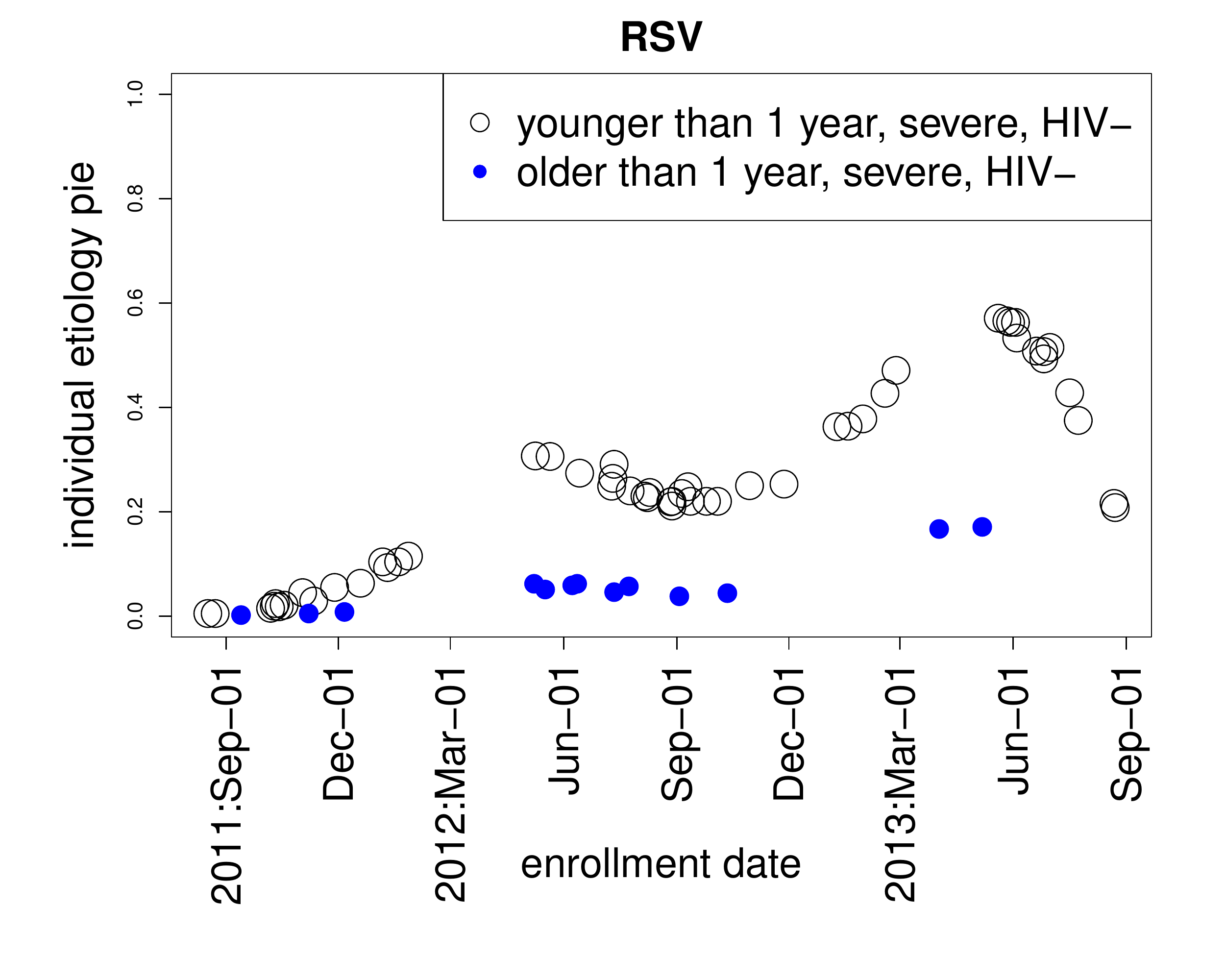}
\label{fig:different_individual_etiology_RSV}
}
}
{\subfigure[Cause: {\sf NoS}]{
\includegraphics[width=0.45\linewidth]{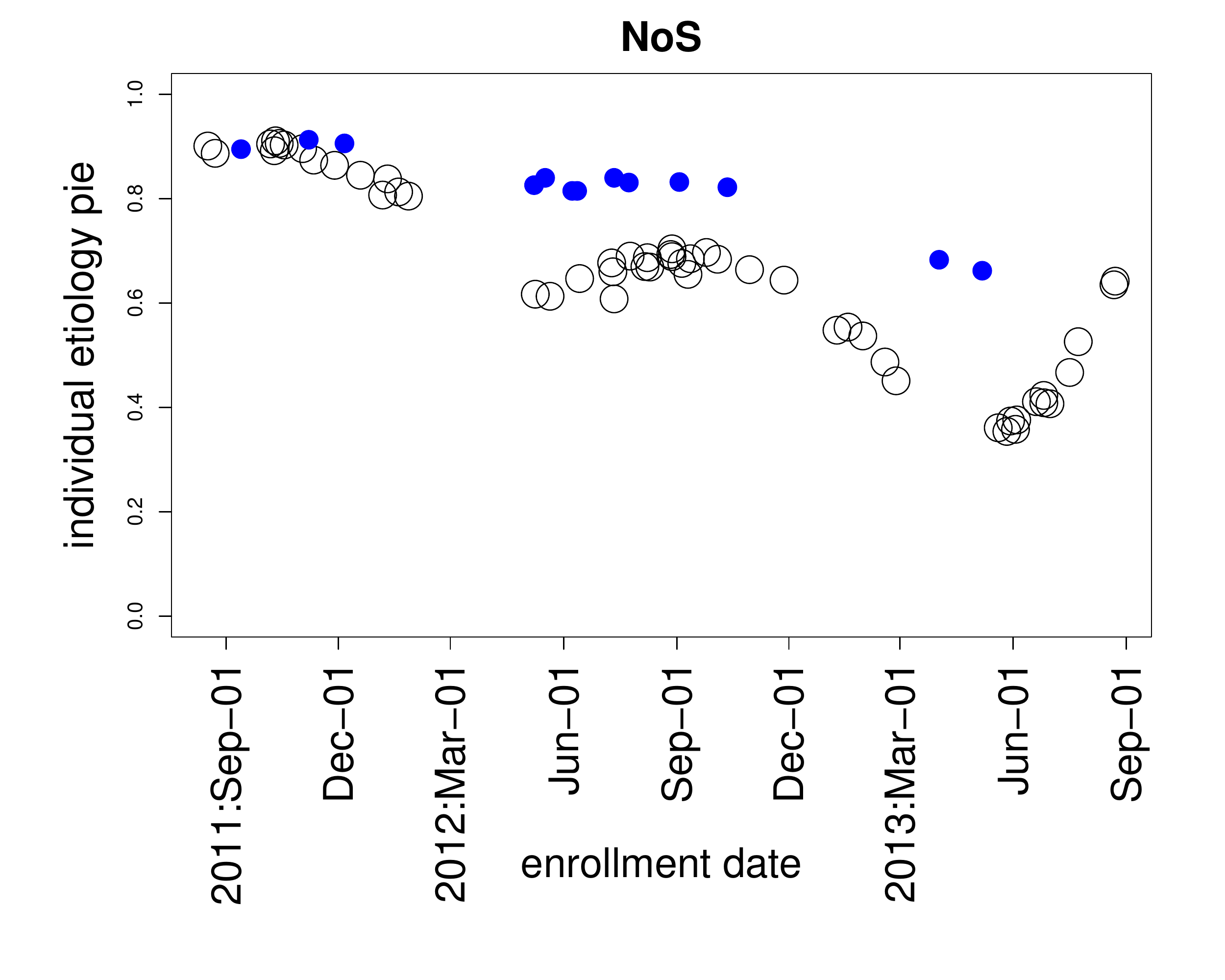}
\label{fig:different_individual_etiology_NoS}
}
}
\addtocounter{figure}{0} 
\caption{\footnotesize Individual etiology fraction estimates for {\sf RSV} (left) and {\sf NoS} (right) differ by age and season among HIV negative and severe pneumonia cases for whom the seven pathogens were \textit{all tested negative} in the nasopharyngeal specimens.}
\end{figure}

\end{document}